\documentclass[aps,pre,reprint,superscriptaddress,longbibliography]{revtex4-2}

\usepackage{graphicx}
\usepackage{amsmath}
\usepackage{amssymb}
\usepackage{bm}
\usepackage{hyperref}
\usepackage{xcolor}
\usepackage{booktabs}

\newtheorem{proposition}{Proposition}

\newenvironment{proof}{\noindent\textit{Proof.}\;}{\hfill$\square$\medskip}

\begin{document}

\title{Exact Conservation Laws of the Lorenz Attractor:\\
Classification and Deterministic Prediction of Lobe-Switching Events}

\author{B.~A.~Toledo}
\email{btoledoc@uchile.cl}
\affiliation{Departamento de F\'{\i}sica, Facultad de Ciencias F\'{\i}sicas y Matem\'aticas,
Universidad de Chile, Santiago, Chile}

\date{\today}

\begin{abstract}
Predicting when a chaotic trajectory will switch between the lobes of
the Lorenz attractor is a long-standing challenge in nonlinear
dynamics. This work shows that algebraic conservation laws,
constructed by augmenting phase space with history-accumulating
auxiliary variables, provide a deterministic solution. Systematic
enumeration identifies eighteen valid invariants in three classes,
each tied to a nullcline of the Lorenz flow, while six candidates
fail, proving that the dynamics constrains which conservation laws are
admissible. One class generates sharp spikes synchronized with
lobe-switching events, achieving $99.2\%$ sensitivity with $0.3\%$
false-positive rate ($\mathrm{AUC} = 0.9995$) as a continuous
Poincar\'e section analogue. The spike amplitude predicts switching
latency via $\Delta t = t_{\min} + C\mathcal{A}^{-n}$ with $R^2 >
0.95$ across all parameter combinations tested. At canonical
parameters $(\sigma, \rho, \beta) = (10, 28, 8/3)$, $n = 2.14 \pm
0.17$ with $R^2 = 0.93$ for individual events; the exponent increases
with $\beta$ and decreases with $\rho$, while the $\sigma$-dependence
is non-monotonic. The latency distribution reveals a topological gap
of width $\Delta t_{\mathrm{gap}} \approx 0.68 \pm 0.01$ for $\rho$
sufficiently above the onset of chaos, explained by the Shilnikov
passage map. Under stochastic perturbations, lobe-sensitive invariants
are ${\sim}\,10^3$ times more robust than their smooth
counterparts. In the Rayleigh-B\'enard convection context, the
auxiliary variables correspond to integrated heat-flux
anomalies. Conservation is verified to $O(10^{-36})$.
\end{abstract}


\maketitle

\section{Introduction}
\label{sec:introduction}

Abrupt transitions between qualitatively different dynamical states
are ubiquitous in nonlinear systems: convective rolls reverse their
circulation, lasers hop between modes, and geophysical flows switch
between competing patterns. In many of these systems the transitions
are recurrent yet individually unpredictable, because the underlying
dynamics are chaotic. Forecasting such events is a central goal of
nonlinear science, with practical implications ranging from weather
prediction to the monitoring of engineering instabilities.

The Lorenz system~\cite{Lorenz1963} provides the paradigmatic setting
for this problem. Derived from a Galerkin truncation of the Boussinesq
equations for Rayleigh-B\'enard convection~\cite{Strogatz2015}, it
describes the interplay between convective circulation (the variable
$x$), horizontal temperature contrast ($y$), and vertical thermal
stratification ($z$). The chaotic attractor has two lobes,
corresponding to opposite senses of fluid circulation, and the
trajectory switches between them in an apparently random
sequence. Predicting when the next ``lobe switch'' will occur has long
been regarded as infeasible, since neighboring initial conditions
diverge exponentially and no conserved quantity was thought to exist
in this dissipative flow.

Two broad strategies have been applied to analogous prediction
problems. Statistical early-warning
indicators~\cite{Scheffer2009,Dakos2012,Lenton2012} detect the
approach to a bifurcation point by monitoring signatures of critical
slowing down, such as rising variance or autocorrelation. These
methods are powerful for systems driven slowly toward a tipping point,
but they do not apply to recurrent events within a stationary chaotic
regime, where no parameter is changing. Symbolic dynamics and
return-map approaches~\cite{Cvitanovic2005,Gilmore1998} classify
transitions post hoc through the sequence of lobe visits, but they do
not provide real-time advance warning of individual events.

A qualitatively different approach was recently opened by the
discovery that the Lorenz system possesses history-dependent dynamical
invariants~\cite{Toledo2026}. By augmenting the three-dimensional
phase space with an auxiliary variable $v(t)$ that accumulates the
trajectory's past, one can construct a quantity $K = P(x,y,z) + v$
that remains exactly constant along every solution. The polynomial $P$
captures the instantaneous configuration, while $v$ compensates for
the dissipation-induced drift, so that their sum is
conserved. Conservation is enforced by requiring the auxiliary
variable to evolve according to a specific rule: $\dot{v} = Q(x,y,z) =
-\dot{P}$, where the ``evolution function'' $Q$ is a fully determined
polynomial of the phase-space coordinates. Although $K$ encodes the
full history, the evolution function $Q$ depends only on the current
state.

A natural question then arises: is there just one such conservation
law, or many? The original construction involved a choice (a specific
ordering and signing of the flow components in the orthogonality
ansatz), and that choice was not unique. In the Mori-Zwanzig formalism
of non-equilibrium statistical
mechanics~\cite{Chorin2000,Zwanzig2001,Givon2004}, projecting a
high-dimensional system onto fewer variables generically produces
memory terms; the auxiliary variable $v(t)$ plays an analogous role by
deterministically resolving what would otherwise be non-Markovian
dynamics. The existence of multiple invariants would therefore reveal
a richer memory structure than a single conserved quantity suggests.

This paper undertakes a systematic exploration of all such
conservation laws. By considering every possible ordering of the four
extended-space coordinates in the constructive ansatz, we enumerate
all candidate invariants and determine which ones survive the
regularity conditions imposed by the Lorenz flow. The result is a
complete family of eighteen valid invariants organized into three
classes, plus a ``null class'' of six candidates that fail because the
Lorenz equations forbid them. This null class is significant: if the
construction were trivially applicable to any polynomial, all
candidates would succeed, and the conservation structure would carry
no information about the specific dynamics.

The central finding, and the primary motivation for this study, is
that the three classes respond to chaotic dynamics in complementary
ways. Class~III invariants are tied to the polynomial
$p_{\mathrm{III}} = xy - \beta z$, which is precisely the right-hand
side of the equation governing vertical thermal stratification
($\dot{z} = xy - \beta z$). This polynomial vanishes on a surface that
intersects the separatrix between the two lobes; consequently, the
Class~III evolution functions $Q_{\mathrm{III}}$ exhibit sharp spikes
whenever the trajectory approaches a lobe switch, while Class~I
evolution functions vary smoothly. The differential response $\Delta S
= Q_{\mathrm{III}} - Q_{\mathrm{I}}$ therefore acts as a geometric
``proximity sensor'' for the separatrix: its amplitude predicts the
time to the next lobe switch via a power law whose exponent is set by
the interplay between the thermal relaxation rate ($\beta$) and the
convective driving ($\rho$).

This precursor signal differs qualitatively from statistical
early-warning indicators. It is deterministic rather than statistical,
instantaneous rather than requiring a sliding time window, and it
operates within a fully developed chaotic regime rather than near a
bifurcation threshold. Its quantitative amplitude-latency relationship
provides not merely a warning that a transition is imminent, but an
estimate of when it will occur.

This paper is organized as follows. Section~\ref{sec:method} reviews
the constructive method and introduces the systematic
enumeration. Section~\ref{sec:family} presents the eighteen invariants
and the null class. Section~\ref{sec:independence} establishes
functional independence. Section~\ref{sec:geometry} develops the
geometric interpretation. Section~\ref{sec:symmetry} analyzes
$\mathbb{Z}_2$ symmetry properties. Section~\ref{sec:numerical}
provides numerical verification, statistical characterization,
topological precursor detection, the origin of the latency gap
(Sec.~\ref{sec:gap}), and the parametric scaling law
(Sec.~\ref{sec:theory}). Section~\ref{sec:discussion} discusses the
non-triviality of the conservation structure
(Sec.~\ref{sec:non-triviality}), the ontological status of the
auxiliary variable (Sec.~\ref{sec:ontology}), the physical
interpretation as an integrated heat-flux anomaly in Rayleigh-B\'enard
convection (Sec.~\ref{sec:experimental}), and the operational reset
protocol.

\section{Systematic Extension of the Constructive Method}
\label{sec:method}

\subsection{Review of the Orthogonality Approach}

The Lorenz system is defined by the three coupled nonlinear ordinary
differential equations
\begin{align}
\dot{x} &= \sigma(y - x), \label{eq:lorenz1}\\
\dot{y} &= x(\rho - z) - y, \label{eq:lorenz2}\\
\dot{z} &= xy - \beta z, \label{eq:lorenz3}
\end{align}
where $\sigma$, $\rho$, and $\beta$ are positive parameters. For the
classical parameter values $\sigma = 10$, $\rho = 28$, and $\beta =
8/3$, the system exhibits chaotic dynamics on the Lorenz
attractor~\cite{Lorenz1963,Eckmann1985,Strogatz2015}.

To construct conserved quantities for this odd-dimensional system, the
phase space is augmented with an auxiliary variable $u$, yielding a
four-dimensional state space $(x, y, z, u)$. A candidate conserved
quantity $C(x, y, z, u)$ must satisfy
\begin{equation}
\frac{dC}{dt} = \nabla C \cdot \mathbf{f} = 0, \label{eq:conservation}
\end{equation}
where $\nabla \equiv (\partial_x, \partial_y, \partial_z,
\partial_u)$, the flow vector is $\mathbf{f} \equiv (\dot{x}, \dot{y},
\dot{z}, \dot{u})$, and $\dot{u} = f(x,y,z)$ is to be determined
self-consistently.

\subsection{Connection to the Mori-Zwanzig Formalism}
\label{sec:mori-zwanzig}

The introduction of the auxiliary variable $u(t)$ admits a natural
interpretation within the framework of non-equilibrium statistical
mechanics. The Mori-Zwanzig projection operator
formalism~\cite{Chorin2000,Zwanzig2001} establishes that when a
high-dimensional system is projected onto a lower-dimensional
subspace, the resulting equations necessarily acquire memory terms:
\begin{equation}
\frac{dA}{dt} = \Omega A + \int_0^t K(t-s) A(s)\, ds + F(t), \label{eq:mori-zwanzig}
\end{equation}
where $K(t)$ is a memory kernel encoding the influence of eliminated
degrees of freedom and $F(t)$ represents a fluctuating force
orthogonal to the resolved variables.

The Lorenz system, derived from a low-order truncation of the
Navier-Stokes equations, represents such a projection. The regularized
auxiliary variable $v_n(t)$, which evolves according to
\begin{equation}
v_n(t) = v_n(0) + \int_0^t Q_n(x(s), y(s), z(s))\, ds, \label{eq:aux-def}
\end{equation}
can be understood as a deterministic resolution of the memory kernel:
it explicitly tracks the accumulated effect of the trajectory's
history that would otherwise manifest as non-Markovian dynamics.

A distinction must be drawn between two different approaches to
treating unresolved degrees of freedom. In the Optimal Prediction
framework of Chorin and collaborators~\cite{Chorin2000}, the
generalized Langevin equation decomposes the dynamics into three
terms: a Markovian self-interaction, a non-Markovian memory integral,
and a fluctuating force $F(t)$ orthogonal to the resolved
variables. First-order optimal prediction is obtained by
\emph{discarding} the memory and noise terms; this approximation is
justified because the conditional expectation $E[F(t) | A] = 0$
vanishes, but the approximation loses accuracy over time as
information about the unresolved degrees of freedom decays. Subsequent
developments~\cite{Givon2004} have placed this framework on rigorous
mathematical footing, establishing convergence results for the
$t$-model and related reduced descriptions of stiff systems. In
contrast, the present construction constitutes a \emph{deterministic
embedding}: by extending the phase space with $v_n(t)$, we absorb the
would-be orthogonal dynamics into a resolved variable, achieving an
exact closure without approximation. Where the Mori-Zwanzig approach
approximates the memory kernel and discards the noise, the invariant
construction resolves both exactly at the cost of introducing an
additional dynamical variable.

A terminological clarification is warranted. The auxiliary variable
$v_n$ is not a hidden degree of freedom in the Mori-Zwanzig sense; it
is a deterministic functional of the resolved trajectory, defined by
$v_n(t) = v_n(0) + \int_0^t Q_n(x(s), y(s), z(s))\, ds$. The analogy
with the memory kernel is structural rather than literal: $v_n$
absorbs the information that would manifest as non-Markovian dynamics
if the four-dimensional extended system were projected back onto three
dimensions, but it does not arise from a projection operator acting on
unresolved modes. The connection to the Mori-Zwanzig formalism is
therefore one of functional role, not of derivation.

Morrison~\cite{Morrison1986} has developed a framework for describing
systems that possess both Hamiltonian and dissipative components,
combining symplectic and metric structures into what are termed
metriplectic systems. While the present construction differs in its
approach, it shares the goal of revealing hidden structure within
dissipative dynamics.

\emph{Physical scaling.} In the original Rayleigh-B\'enard convection
context from which the Lorenz equations derive~\cite{Lorenz1963}, the
state variables represent amplitudes of truncated Fourier modes: $x$
corresponds to the convective velocity mode, while $y$ and $z$
correspond to temperature perturbation modes. The parameters $\sigma$,
$\rho$, and $\beta$ are dimensionless after appropriate scaling by the
thermal diffusivity, cell height, and critical Rayleigh number. In
this dimensionless formulation, the regularized auxiliary variable
$v_n(t)$ inherits a kinematic interpretation as a generalized
potential that accumulates historical effects of the modal
amplitudes. For each regularization class, the polynomial factor
$p_n(x,y,z)$ combines the physical modes in a specific manner:
$p_{\mathrm{I}} = y - x$ couples velocity and temperature
perturbations, $p_{\mathrm{II}} = y + x(z-\rho)$ involves the
departure from criticality, and $p_{\mathrm{III}} = xy - \beta z$
couples to the nonlinear heat flux term.

\subsection{Permutation-Based Polynomial Construction}
\label{sec:permutation-method}

The key insight is that conserved quantities can be constructed
systematically using permutations of the flow components as
\emph{heuristic generators} of polynomial candidates. We adopt a
compact notation for permutations where we identify
\begin{equation}
1 \leftrightarrow x, \quad 2 \leftrightarrow y, \quad 3 \leftrightarrow z, \quad 4 \leftrightarrow u, \label{eq:notation}
\end{equation}
and write permutations as four-digit strings. For example, the permutation 1234 corresponds to the identity ordering $(x, y, z, u)$, while 2134 corresponds to $(y, x, z, u)$.

\emph{The inverse construction method.} Rather than assuming that the
gradient of a conserved quantity equals a permutation of the flow, we
employ the permutation as a generator to synthesize a polynomial
candidate. Given a permutation $\pi$, we construct a candidate scalar
$C$ by summing partial primitives:
\begin{equation}
C(x,y,z,u) = \sum_{i=1}^{4} \int \pi(\mathbf{f})_i \, dq_i, \label{eq:candidate-construction}
\end{equation}
where $\pi(\mathbf{f})_i$ denotes the $i$-th component of the permuted
flow vector and $\mathbf{q} = (x, y, z, u)$.

The resulting candidate $C(x,y,z,u)$ generically takes the form
\begin{equation}
C = P(x,y,z) + p(x,y,z) \cdot u, \label{eq:candidate-form}
\end{equation}
where $P$ is a polynomial and $p$ is a polynomial factor coupling the
auxiliary variable to the physical coordinates.

\emph{Classification by singularity structure.} The polynomial
$p(x,y,z)$ appearing in the coupling term $p \cdot u$ of the invariant
candidate determines the singularity structure. Before regularization,
the auxiliary variable $u$ satisfies an evolution equation of the form
$\dot{u} = F(x,y,z)u + G(x,y,z)$, where both $F$ and $G$ contain
factors of $1/(x-y)$, $1/(y+x(z-\rho))$, or $1/(xy - \beta z)$
depending on the class. The regularization $v = p \cdot u$ removes
these singularities. Remarkably, the 18 valid permutations partition
into exactly three families based on the regularization polynomial:
\begin{itemize}
\item Permutations 1abc yield $p = y - x$ (Class~I)
\item Permutations 2abc yield $p = y + x(z-\rho)$ (Class~II)
\item Permutations 3abc yield $p = xy - \beta z$ (Class~III)
\end{itemize}
This classification by regularization polynomial is the fundamental
organizing principle.

\section{Complete Family of Invariants}
\label{sec:family}

\subsection{Organization by Permutation Class}

The systematic exploration reveals an organizing principle: the first
index in the permutation determines the regularization class. This
leads to a natural partition:
\begin{itemize}
\item \textbf{Class~I} ($K_1$--$K_6$): Permutations 1abc (first coordinate is $x$)
\item \textbf{Class~II} ($K_7$--$K_{12}$): Permutations 2abc (first coordinate is $y$)
\item \textbf{Class~III} ($K_{13}$--$K_{18}$): Permutations 3abc (first coordinate is $z$)
\item \textbf{Null Class}: Permutations 4abc (first coordinate is $u$)
\end{itemize}
The null class consists of permutations where the auxiliary variable
$u$ occupies the first position. These permutations cannot yield valid
invariants because the resulting consistency equations admit only
trivial solutions.

\subsection{Master Table of Invariants}

Table~\ref{tab:master} presents the complete enumeration. Each row
specifies the permutation, the resulting polynomial $P_n(x,y,z)$, the
regularization polynomial $p_n$, and the evolution function $Q_n =
\dot{v}_n$ for the regularized auxiliary variable. The eighteen valid
invariants take the form $K_n = P_n(x,y,z) + v_n$, where conservation
$\dot{K}_n = 0$ requires $\dot{v}_n = -\dot{P}_n = Q_n$. Three
representative evolution functions, one from each class, are given
explicitly in Eqs.~\eqref{eq:Q5}--\eqref{eq:Q13}.

\begin{table*}[htbp]
\caption{Complete family of history-dependent invariants for the
  Lorenz system. Each invariant $K_n = P_n + v_n$ is specified by its
  polynomial part $P_n(x,y,z)$ and evolution function $Q_n =
  \dot{v}_n$. The 24 permutations partition into three valid classes
  (18 invariants) and one null class (6 permutations). Within each
  class, the six permutations of the remaining three indices yield
  distinct polynomial structures sharing a common regularization
  polynomial $p_n$, shown in the class header. Representative
  evolution functions $Q_5$, $Q_{11}$, and $Q_{13}$ are given in
  Eqs.~\eqref{eq:Q5}--\eqref{eq:Q13}.}
\label{tab:master}
\begin{ruledtabular}
\begin{tabular}{ccll}
$n$ & Perm & $P_n(x,y,z)$ & $Q_n(x,y,z)$\\
\hline
\multicolumn{4}{c}{\textbf{Class~I}: $p_{\mathrm{I}} = y - x$}\\
1 & 1234 & $xy - \frac{1}{2}xy^2 + \frac{1}{2}x^2z + \beta yz - \frac{\rho}{2}x^2$ & $-\dot{P}_1$\\
2 & 1243 & $-\frac{1}{2}x^2y + \frac{1}{2}y^2 + xyz + \beta xz - \rho xy$ & $-\dot{P}_2$\\
3 & 1324 & $-\frac{1}{2}xy^2 + yz + \frac{1}{2}xz^2 + \beta yz - \rho xz$ & $-\dot{P}_3$\\
4 & 1342 & $-\frac{1}{2}x^2y + yz + \frac{1}{2}xz^2 + \beta xz - \rho xz$ & $-\dot{P}_4$\\
5 & 1423 & $\frac{1}{2}y^2 + \frac{\beta}{2}z^2 - \rho xy$ & Eq.~\eqref{eq:Q5}\\
6 & 1432 & $xy + \frac{1}{2}x^2z - xyz + \frac{\beta}{2}z^2 - \frac{\rho}{2}x^2$ & $-\dot{P}_6$\vspace{1mm}\\
\hline
\multicolumn{4}{c}{\textbf{Class~II}: $p_{\mathrm{II}} = y + x(z-\rho)$}\\
7 & 2134 & $-\frac{1}{2}xy^2 + \beta yz + \frac{\sigma}{2}x^2 - \sigma xy$ & $-\dot{P}_7$\\
8 & 2143 & $-\frac{1}{2}x^2y + \beta xz + \sigma xy - \frac{\sigma}{2}y^2$ & $-\dot{P}_8$\\
9 & 2314 & $-\frac{1}{2}xy^2 + \beta yz + \sigma xz - \sigma yz$ & $-\dot{P}_9$\\
10 & 2341 & $-\frac{1}{2}x^2y + \beta xz + \sigma xz - \sigma yz$ & $-\dot{P}_{10}$\\
11 & 2413 & $-xyz + \frac{\beta}{2}z^2 + \sigma xy - \frac{\sigma}{2}y^2$ & Eq.~\eqref{eq:Q11}\\
12 & 2431 & $-xyz + \frac{\beta}{2}z^2 + \frac{\sigma}{2}x^2 - \sigma xy$ & $-\dot{P}_{12}$\vspace{1mm}\\
\hline
\multicolumn{4}{c}{\textbf{Class~III}: $p_{\mathrm{III}} = xy - \beta z$}\\
13 & 3124 & $\frac{1}{2}y^2 + xyz - \rho xy + \frac{\sigma}{2}x^2 - \sigma xy$ & Eq.~\eqref{eq:Q13}\\
14 & 3142 & $xy + \frac{1}{2}x^2z - \frac{\rho}{2}x^2 + \sigma xy - \frac{\sigma}{2}y^2$ & $-\dot{P}_{14}$\\
15 & 3214 & $\frac{1}{2}y^2 + xyz - \rho xy + \sigma xz - \sigma yz$ & $-\dot{P}_{15}$\\
16 & 3241 & $xy + \frac{1}{2}x^2z - \frac{\rho}{2}x^2 + \sigma xz - \sigma yz$ & $-\dot{P}_{16}$\\
17 & 3412 & $yz + \frac{1}{2}xz^2 - \rho xz + \sigma xy - \frac{\sigma}{2}y^2$ & $-\dot{P}_{17}$\\
18 & 3421 & $yz + \frac{1}{2}xz^2 - \rho xz + \frac{\sigma}{2}x^2 - \sigma xy$ & $-\dot{P}_{18}$\vspace{1mm}\\
\hline
\multicolumn{4}{c}{\textbf{Null Class}: Permutations 4abc yield trivial solutions only}\\
\end{tabular}
\end{ruledtabular}
\end{table*}

The three representative evolution functions, expressed in terms of
the regularization polynomials $p_{\mathrm{I}} = y - x$,
$p_{\mathrm{II}} = y + x(z-\rho)$, and $p_{\mathrm{III}} = xy - \beta
z$, are:

\begin{align}
  Q_5 &= \bigl[z(\beta-1) + \rho(2+\sigma)\bigr] p_{\mathrm{III}} - (1+\rho\sigma) p_{\mathrm{I}}^2 \notag\\
  &\quad - 2(1+\rho\sigma) x\, p_{\mathrm{I}} \notag\\
&\quad + x^2\bigl[\rho(z-\rho) - 1 - \rho\sigma\bigr] - \beta^2 z^2 + \beta\rho(2+\sigma) z, \label{eq:Q5}\\[6pt]
Q_{11} &= -p_{\mathrm{III}}^2 + \bigl[z(1+2\sigma) - \sigma(1+\rho+\sigma)\bigr] p_{\mathrm{III}} \notag\\
&\quad + \sigma(1+\sigma - z) p_{\mathrm{I}}^2 + 2\sigma(1+\sigma - z) x\, p_{\mathrm{I}} \notag\\
&\quad + x^2\bigl[z^2 - (\rho+2\sigma)z + \sigma(\rho+1+\sigma)\bigr] \notag\\
&\quad + \beta(1+2\sigma) z^2 - \beta\sigma(1+\rho+\sigma) z, \label{eq:Q11}\\[6pt]
Q_{13} &= p_{\mathrm{III}}^2 + \bigl[z(\beta-2-\sigma) + 2\rho + \sigma(1+\rho+2\sigma)\bigr] p_{\mathrm{III}} \notag\\
&\quad + (\sigma z - \sigma\rho - \sigma^2 - 1) p_{\mathrm{I}}^2 + 2(\sigma z - \sigma\rho - \sigma^2 - 1) x\, p_{\mathrm{I}} \notag\\
&\quad + x^2\bigl[-z^2 + 2z(\rho+\sigma) - (\rho^2+2\sigma\rho+2\sigma^2+1)\bigr] \notag\\
&\quad - \beta(2+\sigma) z^2 + \beta\bigl[2\rho + \sigma(1+\rho+2\sigma)\bigr] z. \label{eq:Q13}
\end{align}

These expressions reveal the polynomial complexity underlying the
evolution functions. Each $Q_n$ is a polynomial of degree at most four
in the phase-space coordinates, with coefficients that depend on the
system parameters $(\sigma, \rho, \beta)$. The appearance of
$p_{\mathrm{I}}^2$, $p_{\mathrm{II}}^2$, and $p_{\mathrm{III}}^2$
terms explains the enhanced sensitivity near nullclines: when $p_n \to
0$, the quadratic terms dominate the local variation of $Q_n$,
producing the sharp gradients that serve as topological proximity
sensors.

\subsection{Physical Interpretation of the Regularization Polynomials}

The three regularization polynomials possess clear physical
interpretations within the context of the Lorenz equations. The
Class~I polynomial $p_{\mathrm{I}} = y - x$ vanishes on the manifold
where convective velocity equals horizontal temperature
difference. The Class~II polynomial $p_{\mathrm{II}} = y + x(z -
\rho)$ is precisely the right-hand side of Eq.~\eqref{eq:lorenz2} with
opposite sign, vanishing at the $y$-nullcline where $\dot{y} = 0$. The
Class~III polynomial $p_{\mathrm{III}} = xy - \beta z$ is the
right-hand side of Eq.~\eqref{eq:lorenz3}, vanishing at the
$z$-nullcline where $\dot{z} = 0$. This connection to the nullcline
structure of the Lorenz flow establishes that the regularization
classes encode geometric features of the dynamics.

It is interesting to note that the multiplicative structure $p(x,y,z)
\cdot u_n$ implies a ``gating'' mechanism: when the polynomial
$p(x,y,z)$ vanishes on nullclines, the memory's contribution to the
invariant is momentarily suppressed, regardless of the accumulated
history encoded in $u_n$. This gating mechanism may explain the
topological robustness observed in the statistical analysis, as the
memory coupling is automatically reduced precisely where the flow
undergoes critical transitions.

\subsection{Proof of the Null Class}
\label{sec:null-class}

The six permutations beginning with $u$ (4123, 4132, 4213, 4231, 4312,
4321) cannot produce valid invariants. The failure is not accidental
but reflects a fundamental constraint.

\begin{proposition}
Permutations of the form 4abc yield only trivial solutions $C =
\mathrm{const}$.
\end{proposition}

\begin{proof}
When $u$ occupies the first position, the candidate construction
assigns
\begin{equation}
\frac{\partial C}{\partial x} = \dot{u} = f(x,y,z),
\end{equation}
where $f$ is to be determined. The ansatz also assigns derivatives
$\partial_y C$, $\partial_z C$, and $\partial_u C$ to permuted flow
components. Consistency requires that the mixed partial derivatives
commute (Schwarz integrability conditions).

The condition $\partial_y(\partial_x C) = \partial_x(\partial_y C)$ requires
\begin{equation}
\frac{\partial f}{\partial y} = \frac{\partial}{\partial x}[\sigma(y - x)] = -\sigma,
\end{equation}
implying $f = -\sigma y + g(x,z)$ for some function $g$. The condition $\partial_z(\partial_x C) = \partial_x(\partial_z C)$ then requires
\begin{equation}
\frac{\partial g}{\partial z} = \frac{\partial}{\partial x}[x(\rho - z) - y] = \rho - z,
\end{equation}
yielding $g = (\rho - z)z + h(x)$ for some function $h$. Finally, the condition $\partial_u(\partial_x C) = \partial_x(\partial_u C)$ requires
\begin{equation}
0 = \frac{\partial}{\partial x}[xy - \beta z] = y.
\end{equation}
This is a contradiction for generic trajectories where $y \neq
0$. Therefore, no smooth function $C$ satisfying the ansatz exists
throughout phase space, and only the trivial solution $C =
\mathrm{const}$ is admissible. The same argument, with appropriate
permutations of coordinates, applies to all six 4abc permutations.
\end{proof}

The null class embodies a fundamental asymmetry in the construction:
the auxiliary variable must \emph{respond to} the physical dynamics,
not \emph{drive} them. Placing $u$ in the leading position inverts
this relationship, creating an inconsistency that admits only trivial
resolution.

\section{Functional Independence}
\label{sec:independence}

\subsection{Independence Structure}

The eighteen invariants exhibit a constrained independence
structure. While each $K_n$ is a distinct mathematical object with its
own polynomial $P_n$ and evolution function $Q_n$, they are not all
\emph{functionally independent} in the sense that knowing three of
them (one from each class) determines the values of all others along a
given trajectory. The redundancy arises because invariants within the
same regularization class share a common regularization polynomial and
differ only in their polynomial parts $P_n$, which are functions of
the same three physical coordinates.

\begin{proposition}
Within each regularization class, the difference between any two
invariants depends only on the physical coordinates:
\begin{equation}
K_i - K_j = P_i(x,y,z) - P_j(x,y,z) + (v_i - v_j).
\end{equation}
Since both $K_i$ and $K_j$ are constant along trajectories, so is
their difference. But conservation of $P_i - P_j + (v_i - v_j)$ with
$v_i - v_j$ depending only on the trajectory history implies that the
combination $v_i - v_j$ is determined by $P_j - P_i$ up to an additive
constant set by initial conditions.
\end{proposition}

\emph{Practical implication.} Although 18 invariants exist formally,
only three are functionally independent when the regularized auxiliary
variables are related through the constraint structure. A natural
choice of representatives is the \emph{Triad}: one invariant from each
class, for example $(K_5, K_{11}, K_{13})$. The informational content
of the full invariant family is therefore fundamentally
three-dimensional: the 18 invariants span a three-dimensional space of
independent conserved quantities, and the remaining 15 provide
algebraically dependent combinations.

This redundancy is nonetheless physically valuable for three
reasons. First, the within-class relations $v_i - v_j = (P_j - P_i) +
\mathrm{const}$ provide exact internal consistency checks: any
violation signals numerical error or model breakdown. Second, the
availability of six invariants per class allows optimization of the
Triad representative for specific applications; for instance, $K_5$
(the simplest Class~I polynomial) is computationally efficient for
long-time integration, while more complex polynomials may provide
better signal-to-noise ratios in specific dynamical regimes. Third,
the differential response between classes (not within them) is the
basis for the precursor detector $\Delta S = Q_{\mathrm{III}} -
Q_{\mathrm{I}}$; the redundancy within each class plays no role in the
detection mechanism but permits cross-validation of the detected
events against independent algebraic constraints.

\subsection{Invariant-Specific Auxiliary Variables}
\label{sec:auxiliary-specificity}

A key distinction from the original formulation concerns both the
status of the auxiliary variable and the sign convention employed. In
Ref.~\cite{Toledo2026}, a single history-dependent invariant $K$ was
constructed using an orthogonality ansatz of the form $\nabla C =
(-\dot{y}, -\dot{u}, \dot{z}, -\dot{x})$, which assigns independent
signs $\epsilon_i \in \{+1, -1\}$ to each permuted flow
component. This represents one element of a larger family: if we allow
each of the four components to carry an independent sign, the total
number of distinct ans\"atze becomes $24 \times 2^4/2 = 192$ (the
factor of $1/2$ accounts for the global gauge symmetry $K \to
-K$). The systematic exploration presented here restricts attention to
the 24 ans\"atze with uniform global sign, i.e., $\nabla C =
\pm\pi(\dot{x}, \dot{y}, \dot{z}, \dot{u})$.

A crucial observation is that despite belonging to different
subfamilies (mixed-sign versus uniform-sign), the invariant $K$ from
Ref.~\cite{Toledo2026} and the 18 invariants $K_1$--$K_{18}$ share the
same regularization structure: the singularity-removing polynomials
$p_{\mathrm{I}} = y - x$, $p_{\mathrm{II}} = y + x(z-\rho)$, and
$p_{\mathrm{III}} = xy - \beta z$ appear in both constructions. This
shared regularization suggests that the class structure, determined by
the first index in the permutation, is more fundamental than the
specific sign pattern. In the present classification, the original
invariant corresponds most closely to $K_5$ (Class~I with
regularization polynomial $p_{\mathrm{I}} = y - x$).

More precisely: each invariant $K_n$ is associated with a specific
evolution function $Q_n(x,y,z)$, and the regularized auxiliary
variable $v_n$ satisfies $\dot{v}_n = Q_n$. Two invariants $K_i$ and
$K_j$ from different classes have $Q_i \neq Q_j$, so their auxiliary
variables evolve differently even along the same trajectory.

This clarifies the mathematical structure: the 18 invariants do not
share a common auxiliary variable but rather define 18 different
extensions of the three-dimensional Lorenz phase space into four
dimensions. The extensions are related through the constraint that the
physical projection $(x,y,z)$ satisfies the same Lorenz equations in
all cases. The systematic exploration of invariants with independent
component signs constitutes a natural extension of this work.

\section{Geometric Interpretation}
\label{sec:geometry}

\subsection{Foliation of Extended Phase Space}

Each invariant $K_n = P_n(x,y,z) + v_n = c_n$ defines a
three-dimensional hypersurface in the four-dimensional extended phase
space $(x, y, z, v_n)$. The collection of all such hypersurfaces,
parameterized by $c_n \in \mathbb{R}$, constitutes a foliation.

\emph{Projection to physical space.} For a given constant $c_n$, the
constraint $v_n = c_n - P_n(x,y,z)$ determines the auxiliary variable
as a function of position. The physical trajectory $(x(t), y(t),
z(t))$ is accompanied by a unique ``shadow'' $v_n(t) = c_n - P_n(x(t),
y(t), z(t))$ that maintains conservation.

\emph{Intersection structure.} A trajectory in the extended space lies
on the intersection of three independent hypersurfaces (one from each
class). For the Triad $(K_5, K_{11}, K_{13})$ with constants $(c_1,
c_2, c_3)$, the trajectory is confined to a one-dimensional curve in
$(x, y, z, v_5, v_{11}, v_{13})$ space.

\subsection{Characterization of Unstable Periodic Orbits}

Unstable periodic orbits (UPOs) embedded in the strange attractor
acquire a natural characterization through the invariants. For a UPO
with period $T$, the trajectory returns to its initial point: $(x(T),
y(T), z(T)) = (x(0), y(0), z(0))$.

The polynomial parts $P_n$ therefore satisfy $P_n(T) =
P_n(0)$. Conservation $K_n = P_n + v_n = c_n$ then requires
\begin{equation}
v_n(T) - v_n(0) = 0,
\end{equation}
meaning the cycle-integrated evolution function vanishes:
\begin{equation}
\oint_{\mathrm{UPO}} Q_n \, dt = 0. \label{eq:upo-constraint}
\end{equation}

This provides a geometric characterization: UPOs are trajectories for
which all auxiliary variables return to their initial values after one
period. The Triad constants $(c_1, c_2, c_3)$ serve as intrinsic
labels distinguishing different periodic orbits, analogous to action
variables in integrable systems~\cite{Cvitanovic2005}.

\subsection{Visualization of Constraint Surfaces}

Invariants from different regularization classes ``observe'' the
trajectory through fundamentally different geometric lenses: Class~I
responds to the $y-x$ nullcline structure, while Class~III responds to
the $z$-nullcline and its connection to lobe-switching events. This
geometric distinction drives the statistical divergence observed in
the chaotic regime (Sec.~\ref{sec:numerical}).

\begin{figure}[!ht]
  \centering 
  \includegraphics[width=0.95\linewidth]{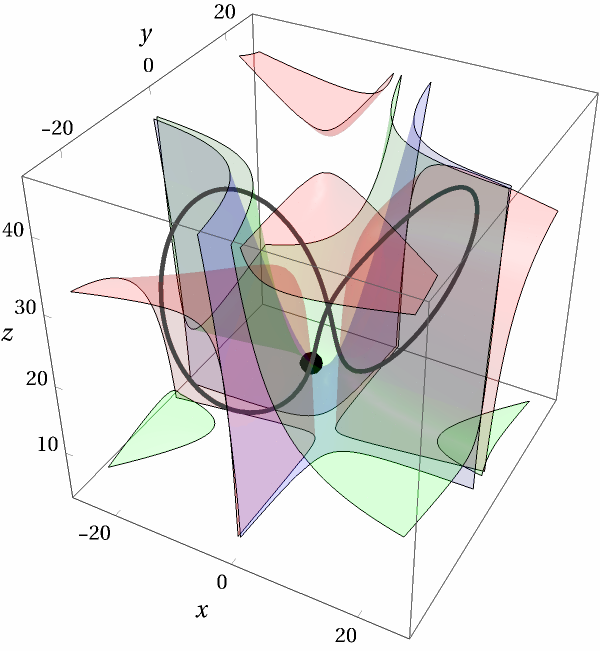}\\
  \includegraphics[width=0.95\linewidth]{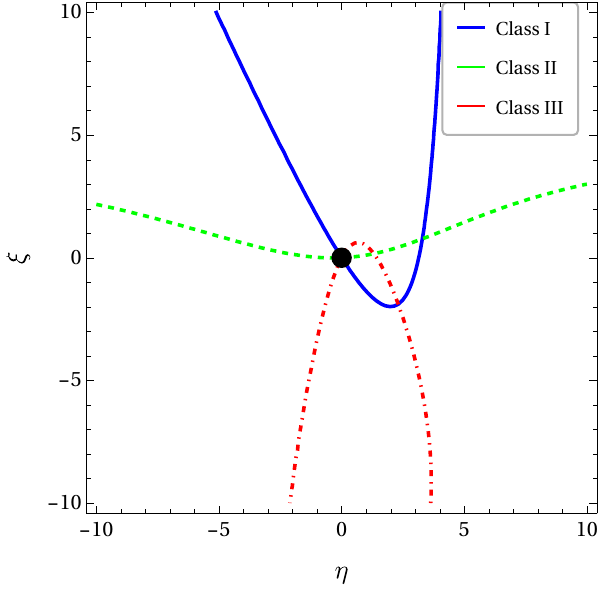}
  \caption{\textbf{Geometric structure of the regularization classes.}
    (Left)~Three-dimensional representation of the level surfaces of
    the invariant polynomials in the Lorenz phase space: $P_5$ (blue),
    $P_{11}$ (green), and $P_{13}$ (red). The unstable periodic orbit
    (dark curve) lies on the simultaneous intersection of these
    constraint surfaces. The black dot marks a representative point on
    the UPO. (Right)~Two-dimensional cross-section in local
    coordinates $(\eta, \xi)$ on an optimized Poincar\'e plane showing
    the topological relationship between the three regularization
    classes: $K_5 = c_1$ (Class~I, blue solid), $K_{11} = c_2$
    (Class~II, green dashed), and $K_{13} = c_3$ (Class~III, red
    dot-dashed). The intersection point corresponds to configurations
    where all three polynomial constraints are satisfied
    simultaneously on the UPO. Parameters: $\sigma = 10$, $\rho = 28$,
    $\beta = 8/3$.}
  \label{fig:surfaces}
\end{figure}

Figure~\ref{fig:surfaces} displays constraint surfaces for the
canonical Triad $\mathcal{T} = \{K_5, K_{11}, K_{13}\}$ anchored to an
unstable periodic orbit (UPO). The three-dimensional visualization
(upper panel) depicts the level surfaces of the invariant polynomials
in the Lorenz phase space: $P_5(x,y,z) = c_1$ (Class~I, blue),
$P_{11}(x,y,z) = c_2$ (Class~II, green), and $P_{13}(x,y,z) = c_3$
(Class~III, red), where the constants $c_n$ are determined by the UPO
geometry. The unstable periodic orbit (dark curve) illustrates the
system's evolution relative to these intersecting constraint surfaces.

Lower panel presents a two-dimensional cross-section on an optimized
Poincar\'e plane. The local coordinates $(\eta, \xi)$ are defined by
$\mathbf{P}(\eta, \xi) = \mathbf{P}_0 + \eta \, \mathbf{\hat{u}} + \xi
\, \mathbf{\hat{w}}$, where $\mathbf{P}_0$ is a point on the unstable
periodic orbit and $\{\mathbf{\hat{u}}, \mathbf{\hat{w}}\}$ form an
orthonormal basis of a plane whose orientation was numerically
optimized to maximize the angular separation between the three level
curves. The curves meet at a single point (the UPO crossing, the same
point in the upper panel), confirming that all three constraints are
simultaneously satisfied.

\section{Symmetry Properties}
\label{sec:symmetry}

\subsection{$\mathbb{Z}_2$ Symmetry of the Lorenz System}

The Lorenz equations are equivariant under the discrete symmetry
\begin{equation}
\mathcal{R}: (x, y, z) \mapsto (-x, -y, z), \label{eq:z2-symmetry}
\end{equation}
corresponding to a $180°$ rotation about the $z$-axis. This symmetry
permutes the two lobes of the attractor and exchanges the symmetric
fixed points $C_+ \leftrightarrow C_-$ located at
$(\pm\sqrt{\beta(\rho-1)}, \pm\sqrt{\beta(\rho-1)}, \rho-1)$.

\subsection{Transformation of Invariants}

Under $\mathcal{R}$, the polynomial parts transform according to their
monomial content:
\begin{itemize}
\item Terms with even total degree in $(x,y)$ are invariant: $z^2$,
  $x^2$, $y^2$, $xy$, $x^2z$, $\ldots$
\item Terms with odd total degree change sign: $x$, $y$, $xz$, $yz$,
  $\ldots$
\end{itemize}

Analysis of Table~\ref{tab:master} reveals that:
\begin{itemize}
\item $P_5 = \frac{1}{2}y^2 + \frac{\beta}{2}z^2 - \rho xy$ is
  $\mathcal{R}$-invariant (even in $x,y$).
\item $P_{11} = -xyz + \frac{\beta}{2}z^2 + \sigma xy -
  \frac{\sigma}{2}y^2$ is $\mathcal{R}$-invariant, since all terms
  have even degree in $(x,y)$.
\item $P_{13}$ contains mixed terms and transforms non-trivially.
\end{itemize}

The regularization polynomials transform as:
\begin{align}
\mathcal{R}(y - x) &= -y - (-x) = -(y - x), \\
\mathcal{R}(y + x(z-\rho)) &= -y + (-x)(z-\rho) = -(y + x(z-\rho)), \\
\mathcal{R}(xy - \beta z) &= (-x)(-y) - \beta z = xy - \beta z.
\end{align}

\emph{Complete parity analysis.} The transformation properties of the
full invariants $K_n = P_n + v_n$ depend on both the polynomial part
$P_n$ and the regularized auxiliary variable $v_n = p_n \cdot
u$. Since the auxiliary variable $u$ satisfies a first-order ODE
driven by the Lorenz flow, and the initial condition is fixed at $u(0)
= 0$ (a gauge choice that respects the $\mathbb{Z}_2$ symmetry), the
parity of $v_n$ is determined by the parity of $p_n$. The Class~III
regularization polynomial $p_{\mathrm{III}} = xy - \beta z$ is
$\mathcal{R}$-invariant (even), while the Class~I and~II
regularization polynomials are $\mathcal{R}$-odd. However, the overall
parity of each invariant $K_n$ also depends on the parity of $P_n$: an
invariant is $\mathcal{R}$-even only if both $P_n$ and $p_n$ have
matching parities that combine to yield an even transformation. Within
each class, the individual invariants may have different parities
depending on their specific polynomial structure.

\subsection{Implications for Symmetry-Related Orbits}

For a trajectory $\gamma(t)$ and its symmetric image
$\mathcal{R}\gamma(t)$, the Triad constants satisfy specific
relationships determined by the transformation properties
above. Symmetric periodic orbits (those invariant under $\mathcal{R}$)
have Triad constants constrained by these symmetry relations,
providing additional structure for orbit classification.

\section{Numerical Verification and Statistical Characterization}
\label{sec:numerical}

\subsection{High-Precision Validation}

Conservation was verified using extended-precision arithmetic (80
decimal digits) to eliminate concerns about numerical
drift. Integration employed a stiffness-switching adaptive
algorithm~\cite{Toledo2026} that selects between explicit and implicit
solvers based on local stiffness detection, with accuracy and
precision goals set to 70 digits, maintaining local error below
$10^{-70}$.

Table~\ref{tab:triad} presents conservation errors for the canonical
Triad after one UPO period. Errors of order $10^{-36}$--$10^{-38}$
confirm conservation to numerical precision. With 80-digit working
precision, the machine epsilon is $\sim 10^{-80}$; the observed errors
of $O(10^{-37})$ reflect normal accumulation of roundoff during
integration, confirming that the conservation laws are satisfied to
the limits of numerical precision.

\begin{table}[t]
\caption{Conservation errors $|K(T) - K(0)|$ for Triad invariants
  after one UPO period ($T \approx 1.56$). Calculations performed with
  80-digit working precision.}
\label{tab:triad}
\begin{ruledtabular}
\begin{tabular}{ccc}
Invariant & Class & Error\\
\hline
$K_5$ & I & $3.45 \times 10^{-36}$\\
$K_{11}$ & II & $9.35 \times 10^{-38}$\\
$K_{13}$ & III & $3.68 \times 10^{-38}$\\
\end{tabular}
\end{ruledtabular}
\end{table}

\subsection{Statistical Analysis of Auxiliary Evolution Functions}

We examine whether invariants from different regularization classes
exhibit distinct dynamical signatures. For each invariant $K_n =
P_n(x,y,z) + v_n$, the evolution function
\begin{equation}
Q_n(x,y,z) \equiv \dot{v}_n = -\frac{dP_n}{dt} \label{eq:Qn-def}
\end{equation}
represents the instantaneous rate at which $v_n$ evolves to maintain
conservation. Statistics were computed from $N = 4 \times 10^6$
trajectory samples (integration time $T = 400$ after transient
removal, sampling interval $\Delta t = 10^{-4}$).

\emph{Pre-chaotic regime ($\rho = 23$).} Both classes yield
statistically similar distributions
(Fig.~\ref{fig:statistics}a). Kurtosis values $\kappa_{\mathrm{I}}
\approx 28.1$ and $\kappa_{\mathrm{III}} \approx 28.3$ are effectively
identical. This baseline establishes that there is no intrinsic bias
in the Class~III formulation; the statistical equivalence confirms
that both classes respond identically to the simple spiral dynamics
without lobe-switching.

\emph{Chaotic regime ($\rho = 28$).} A clear structural divergence
emerges (Fig.~\ref{fig:statistics}b). The standardized distributions
reveal different tail structures:

\emph{Differential intermittency.} Class~III exhibits substantially
higher kurtosis ($\kappa_{\mathrm{III}} \approx 15.8$) compared to
Class~I ($\kappa_{\mathrm{I}} \approx 8.2$), a 93\% increase. The
heavy tails extending beyond $\pm 5\sigma$ indicate that Class~III
conservation is subject to sporadic large-amplitude events
(``spikes'') punctuating quiescent intervals.

\emph{Asymmetry structure.} Both classes display pronounced negative
skewness in the chaotic regime, with Class~III exhibiting even
stronger asymmetry ($S_{\mathrm{III}} \approx -2.73$) than Class~I
($S_{\mathrm{I}} \approx -1.66$). This indicates that extreme negative
excursions dominate the fluctuation statistics for both classes, but
the effect is amplified in Class~III due to the rapid sign changes of
the $xy$ term during separatrix crossings. The enhanced negative
skewness of Class~III reflects the geometric asymmetry of
lobe-switching events: trajectories accelerate sharply as they
approach the separatrix from one direction.

\emph{Interpretation.} The elevated kurtosis of Class~III correlates
with lobe-switching events. At separatrix crossings,
$P_{\mathrm{III}}$ varies rapidly due to its $xy$-terms, forcing rapid
adjustments in the Class~III evolution function. Class~I quantities
lack this geometric sensitivity due to the smoothing effect of the
$z^2$ term in $P_5 = y^2/2 + \beta z^2/2 - xy\rho$.

\emph{Pre-chaotic versus chaotic kurtosis.} A noteworthy feature is
that the pre-chaotic regime ($\rho = 23$) exhibits substantially
higher kurtosis ($\kappa \approx 28.2$) than the chaotic regime
($\kappa_{\mathrm{I}} \approx 8.2$). This counterintuitive result
reflects a fundamental difference in statistical structure. At $\rho =
23$, the system operates near the subcritical Hopf bifurcation where
trajectories exhibit complex transient dynamics: long intervals of
quiescence near local attractors are punctuated by abrupt
escapes. These rare but intense bursts generate distributions with
pronounced heavy tails, yielding elevated kurtosis. In contrast, fully
developed chaos at $\rho = 28$ produces more ergodic and mixing
dynamics, with fluctuations distributed more uniformly across time.

\begin{table}[t]
\caption{Statistical characterization of evolution functions $Q_n(t)$
  for Class~I ($K_5$) and Class~III ($K_{13}$). Kurtosis $\kappa$
  measures tail weight (Gaussian: $\kappa = 3$); skewness $S$ measures
  asymmetry. Statistics computed from $N = 4 \times 10^6$ samples over
  $T = 400$ time units.}
\label{tab:statistics}
\begin{ruledtabular}
\begin{tabular}{ccccc}
 & \multicolumn{2}{c}{$\rho = 23$ (pre-chaotic)} & \multicolumn{2}{c}{$\rho = 28$ (chaotic)}\\
Statistic & $Q_{\mathrm{I}}$ & $Q_{\mathrm{III}}$ & $Q_{\mathrm{I}}$ & $Q_{\mathrm{III}}$\\
\hline
Kurtosis $\kappa$ & 28.1 & 28.3 & 8.2 & 15.8\\
Skewness $S$ & $-0.16$ & $-0.44$ & $-1.66$ & $-2.73$\\
\end{tabular}
\end{ruledtabular}
\end{table}

\subsection{Differential Robustness Under Stochastic Perturbations}

The statistical analysis established that Class~III evolution
functions exhibit violent bursts at lobe-switching events
($\kappa_{\mathrm{III}} \approx 15.8$), while Class~I evolution
functions evolve more smoothly ($\kappa_{\mathrm{I}} \approx
8.2$). This distinction might suggest that Class~III invariants would
be more susceptible to noise-induced degradation. However, stochastic
simulations reveal a striking inversion of this intuition.

\emph{Numerical validation.} To assess robustness, we integrate the
Lorenz system with additive Gaussian white noise,
\begin{equation}
d\mathbf{x} = \mathbf{F}(\mathbf{x})\, dt + \epsilon\, d\mathbf{W},
\end{equation}
where $\mathbf{W}$ is a three-dimensional Wiener process and $\epsilon
= 0.1$ controls the noise amplitude. Under such perturbations, the
exact cancellation ensuring $dK/dt = 0$ fails. The invariant $K$
becomes a stochastic process, and its variance grows with
time. Fixed-step Euler-Maruyama integration with ensemble averaging
yields the diffusivity coefficients:
\begin{align}
D_{K_5} &\equiv \frac{d\,\mathrm{Var}(K_5)}{dt} \approx 5.6 \times
10^5, \label{eq:diff-K5}\\ D_{K_{13}} &\equiv
\frac{d\,\mathrm{Var}(K_{13})}{dt} \approx 6.7 \times
10^2. \label{eq:diff-K13}
\end{align}
The ratio $D_{K_{13}}/D_{K_5} \approx 1.2 \times 10^{-3}$ demonstrates
that Class~III invariants are approximately \emph{three orders of
magnitude more robust} under stochastic perturbations than Class~I
invariants. This factor of $\sim 836$ establishes Class~III quantities
as combinatorially robust observables.

\emph{Physical interpretation.} This counterintuitive result admits a
transparent physical explanation rooted in the distinction between
\emph{metric} and \emph{topological} stability:
\begin{itemize}
\item \emph{Metric instability of Class~I.} The polynomial part of
  $K_5$ contains the term $\frac{1}{2}\dot{y}^2$, which depends
  quadratically on the convective velocity. On the chaotic attractor,
  $\dot{y}$ fluctuates continuously with large amplitude, and small
  perturbations in the trajectory propagate quadratically into the
  invariant. This constitutes a persistent, cumulative source of error
  throughout the entire orbital cycle.

\item \emph{Topological stability of Class~III.} The regularization
  polynomial $p_{\mathrm{III}} = xy - \beta z$ encodes the
  trajectory's position relative to the separatrix. While Class~III
  evolution functions exhibit violent bursts at lobe-switching events
  (Fig.~\ref{fig:statistics}), these bursts are \emph{discrete and
  transient}: between crossings, the invariant evolves smoothly with
  minimal error accumulation. Unless the noise is sufficiently strong
  to induce an erroneous lobe transition (effectively a topological
  error), the Class~III invariants remain stable.
\end{itemize}

\emph{Practical implications.} For applications involving noisy or
experimental data, Class~III invariants provide substantially more
robust observables than Class~I, despite their higher instantaneous
intermittency. The elevated kurtosis of Class~III evolution functions
reflects sensitivity to \emph{discrete topological events}, not
continuous metric degradation. In this sense, Class~III invariants
function as ``topologically protected'' quantities whose conservation
is robust against perturbations that do not alter the symbolic
itinerary. A terminological clarification is warranted: the robustness
described here is \emph{combinatorial} rather than topological in the
sense employed in condensed matter physics, where protection is
guaranteed by a quantized topological invariant (such as a Chern
number or a winding number) whose constancy under continuous
deformations is mathematically exact. The protection observed in the
present context is weaker: it holds for perturbations that preserve
the discrete symbolic itinerary (the ordered sequence of lobe visits),
but can be violated by noise of sufficient amplitude to induce
erroneous lobe transitions. The precise characterization of the
admissible perturbation class and its connection to the topology of
the branched manifold constitutes an open question.

The distinction between high kurtosis (intermittency) and high
variance (accumulated error) is fundamental: the former characterizes
the \emph{temporal distribution} of adjustment events, while the
latter determines the \emph{total drift} under noise. The numerical
results establish that these quantities are not positively correlated;
indeed, the class exhibiting higher intermittency possesses lower
accumulated variance by a factor of $\sim 10^3$.

\subsection{Class~III as a Topological Probe}

The Class~III evolution function $Q_{\mathrm{III}}(t)$ functions as a
\emph{symbolic marker} that signals each addition to the trajectory's
symbolic sequence. In the standard symbolic dynamics
description~\cite{Gilmore1998}, each visit to the left lobe is encoded
as $L$, each visit to the right lobe as $R$. The sequence $\ldots
LLRLLRLR \ldots$ captures the trajectory's itinerary through the
attractor.

The evolution function $Q_{\mathrm{III}}(t)$ provides a continuous
encoding of this discrete dynamics:
\begin{itemize}
\item \emph{Quiescent intervals} ($|Q_{\mathrm{III}}| \ll
  |Q_{\mathrm{III}}|_{\text{max}}$): The trajectory remains within a
  single lobe; no symbols are added.
\item \emph{Spike events} ($|Q_{\mathrm{III}}| \sim |Q_{\mathrm{III}}|_{\text{max}}$): The trajectory crosses the separatrix; a new symbol is appended.
\end{itemize}

This establishes $Q_{\mathrm{III}}(t)$ as a \emph{geometric probe of
topological transitions}: the integrated area under each spike
corresponds to the ``topological contribution'' of that symbol
transition. The Class~III evolution functions thereby encode the
discrete symbolic structure of the chaotic flow, while Class~I
evolution functions probe the continuous dynamics.

\emph{Quantitative validation of the symbolic correspondence.} To
substantiate the claim that $Q_{\mathrm{III}}(t)$ faithfully encodes
the symbolic dynamics, we perform a systematic comparison between the
spike events in $|Q_{13}(t)|$ and the zero-crossings of $x(t)$ that
define the symbolic partition. Over a chaotic trajectory of duration
$T = 2500$ time units at the standard parameters ($\sigma = 10$, $\rho
= 28$, $\beta = 8/3$), we identify all separatrix crossings
(zero-crossings of $x(t)$ with either sign of $\dot{x}$) and all spike
events in $|Q_{13}(t)|$ exceeding a threshold
$\mathcal{A}_{\mathrm{th}}$.

The threshold is defined as $\mathcal{A}_{\mathrm{th}} =
\mu_{|Q_{13}|} + k\,\sigma_{|Q_{13}|}$, where $\mu$ and $\sigma$
denote the mean and standard deviation of $|Q_{13}|$ along the
trajectory, and the multiplier $k$ is chosen to optimize the trade-off
between sensitivity and false positive rate. For each spike event (a
contiguous interval during which $|Q_{13}|$ exceeds
$\mathcal{A}_{\mathrm{th}}$, represented by the time of its maximum),
we search for a separatrix crossing within a temporal window
$[t_{\mathrm{spike}} - \delta, t_{\mathrm{spike}} + \delta]$ with
$\delta = 0.5$ time units. A spike is classified as a \emph{true
positive} if a separatrix crossing falls within this window, a
\emph{false positive} if no crossing is found, and a separatrix
crossing without an associated spike constitutes a \emph{missed
event}.

The analysis identifies $N_{\mathrm{cross}} = 1{,}383$ separatrix
crossings over the full trajectory. Table~\ref{tab:symbolic}
summarizes the detection performance as a function of the threshold
multiplier $k$. The sensitivity (fraction of crossings detected)
remains stable at $99.2\%$ for $k \leq 2.25$, then degrades as the
threshold eliminates genuine but weaker spikes. The false positive
rate decreases monotonically with increasing $k$. The optimal
trade-off is achieved at $k = 2.25$, where 1{,}843 spike events are
detected with a sensitivity of $99.2\%$ and a false positive rate of
$0.3\%$; the area under the receiver operating characteristic curve is
$\mathrm{AUC} = 0.9995$, confirming near-perfect discrimination. Only
11 of the 1{,}383 crossings (0.8\%) lack a corresponding spike above
threshold.

\begin{table}[t]
\caption{Detection performance of $|Q_{13}(t)|$ as a symbolic dynamics
  detector, as a function of the threshold multiplier $k$ in
  $\mathcal{A}_{\mathrm{th}} = \mu + k\sigma$. Sensitivity denotes the
  fraction of separatrix crossings detected; FPR denotes the fraction
  of spike events without a corresponding crossing within $\delta =
  0.5$ time units. The optimal threshold ($k = 2.25$) is highlighted.}
\label{tab:symbolic}
\begin{ruledtabular}
\begin{tabular}{cccc}
$k$ & Spikes & Sensitivity (\%) & FPR (\%)\\
\hline
1.0 & 4{,}115 & 99.2 & 18.7\\
1.5 & 2{,}976 & 99.2 & 17.1\\
2.0 & 2{,}221 & 99.2 & 13.6\\
\textbf{2.25} & \textbf{1{,}843} & \textbf{99.2} & \textbf{0.3}\\
2.5 & 1{,}680 & 98.6 & 0.0\\
3.0 & 1{,}287 & 89.2 & 0.0\\
3.5 & 1{,}013 & 71.4 & 0.0\\
4.0 & 827 & 59.3 & 0.0\\
\end{tabular}
\end{ruledtabular}
\end{table}

These results confirm that the Class~III evolution function provides a
high-fidelity continuous encoding of the discrete symbolic
dynamics. The missed events (0.8\%) correspond to grazing separatrix
crossings where the trajectory barely touches the $x = 0$ plane before
returning to the same lobe; such events produce spikes below threshold
because the near-tangential approach does not activate the full
geometric coupling of $p_{\mathrm{III}} = xy - \beta z$. The residual
false positives (0.3\% at $k = 2.25$) correspond to near-separatrix
approaches where the trajectory enters the geometric ``zone of
influence'' of the separatrix (producing a spike in $|Q_{13}|$)
without completing a full crossing. These are not detection errors but
rather genuine geometric proximity events: the evolution function
responds to the local configuration of the phase-space coordinates
relative to the $z$-nullcline surface, and a trajectory that
approaches $x = 0$ closely will activate this response regardless of
whether the crossing is completed.

\begin{figure}[!ht]
  \centering
  \includegraphics[width=0.95\linewidth]{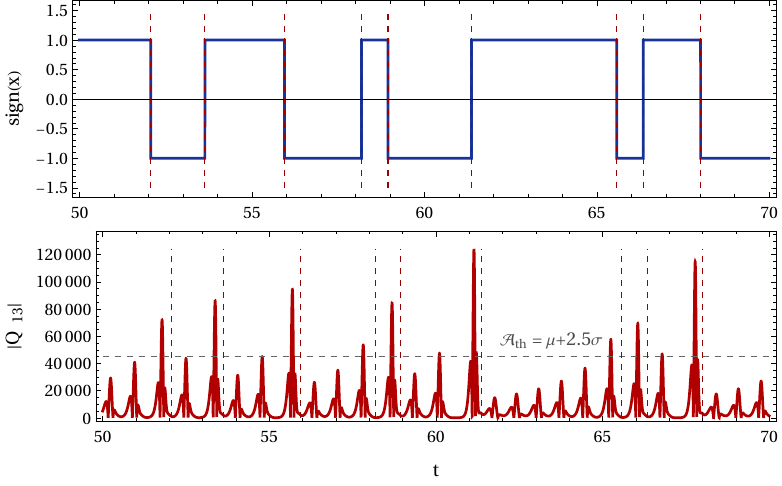}
  \caption{Correspondence between Class~III evolution function and
    symbolic dynamics. Upper panel: symbolic itinerary
    $\mathrm{sign}(x(t))$ over $t \in [50, 70]$ at standard
    parameters. Lower panel: absolute value of the Class~III evolution
    function $|Q_{13}(t)|$ over the same interval, with detection
    threshold $\mathcal{A}_{\mathrm{th}} = \mu + 2.25\sigma$ (dashed
    line). Vertical dotted lines mark separatrix crossings ($x =
    0$). Every transition in the symbolic sequence is accompanied by a
    spike in $|Q_{13}|$ exceeding the threshold, confirming the
    quantitative correspondence between algebraic structure and
    topological dynamics established in Table~\ref{tab:symbolic}.}
  \label{fig:symbolic}
\end{figure}

Figure~\ref{fig:symbolic} illustrates the correspondence over a
representative time interval. The upper panel displays
$\mathrm{sign}(x(t))$, encoding the symbolic itinerary (L for $x < 0$,
R for $x > 0$). The lower panel shows $|Q_{13}(t)|$ with the detection
threshold (horizontal dashed line). Every symbol change in the upper
panel is accompanied by a spike in the lower panel, demonstrating the
functional equivalence between the evolution function and a continuous
Poincar\'e section detector.

\subsection{Topological Precursor Detection}
\label{sec:precursor}

The differential response between Class~III and Class~I evolution
functions provides the basis for a topological precursor detector. We
construct the differential signal
\begin{equation}
\Delta S(t) = Q_{13}(t) - Q_5(t), \label{eq:differential-signal}
\end{equation}
which isolates the topological sensitivity of Class~III from the
continuous dynamics tracked by Class~I.

\emph{Detection protocol.} Define an alert threshold
$\mathcal{A}_{\mathrm{th}}$ based on the amplitude of $\Delta
S(t)$. When $|\Delta S(t)|$ exceeds $\mathcal{A}_{\mathrm{th}}$, a
precursor event is registered. The latency $\Delta t$ is defined as
the time interval between the precursor signal and the subsequent
separatrix crossing (zero-crossing of $x(t)$).

\begin{figure}[!h]
  \includegraphics[width=\linewidth]{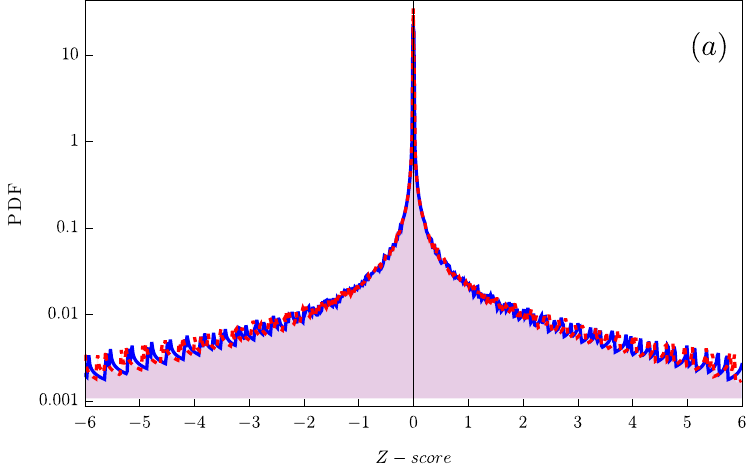}\\
  \includegraphics[width=\linewidth]{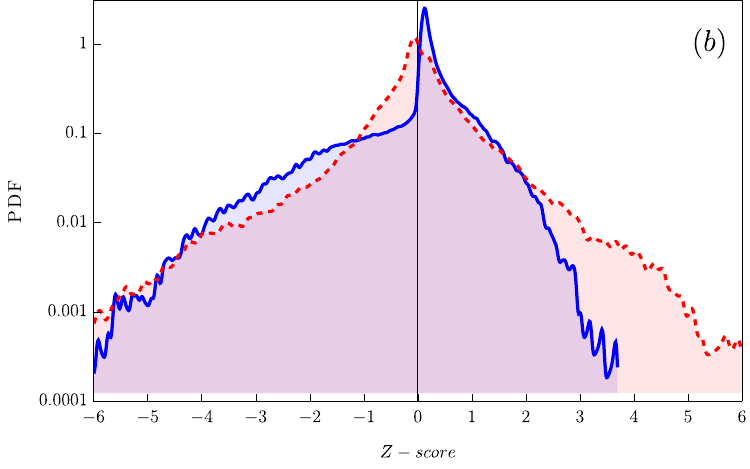}\\
  \includegraphics[width=0.82\linewidth]{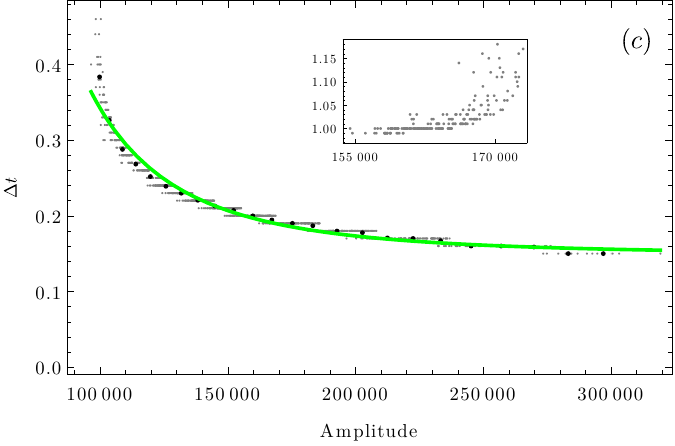}
  \caption{Statistical characterization of evolution functions.
    (a)~Pre-chaotic regime ($\rho = 23$): Class~I and Class~III
    distributions are statistically similar ($\kappa \approx 28.2$).
    (b)~Chaotic regime ($\rho = 28$): Class~III develops heavier tails
    ($\kappa_{\mathrm{III}} \approx 15.8$ versus $\kappa_{\mathrm{I}}
    \approx 8.2$), indicating enhanced intermittency at topological
    transitions.  (c)~Latency $\Delta t$ versus spike amplitude
    $\mathcal{A}$ of the differential signal $\Delta S = Q_{13} - Q_5$
    (gray: individual events; circles: binned averages; curve: fitted
    power law). The gap in $0.45 \leq \Delta t < 0.80$ (standard
    parameters; see Sec.~\ref{sec:gap} for parameter-dependent
    boundaries) reflects a topological sorting mechanism
    (Sec.~\ref{sec:gap}); the exponent $n$ depends on the system
    parameters (Sec.~\ref{sec:theory}).  All panels: $N = 4 \times
    10^6$ samples.}
  \label{fig:statistics}
\end{figure}

\emph{Scaling law.} Figure~\ref{fig:statistics}(c) reveals a
remarkable empirical relationship between the spike amplitude
$\mathcal{A}$ and the prediction latency $\Delta t$. The data are
well-described by a shifted power law:
\begin{equation}
\Delta t = t_{\min} + C \mathcal{A}^{-n}, \label{eq:scaling-law}
\end{equation}
with fitted parameters at canonical parameters $(\sigma, \rho, \beta)
= (10, 28, 8/3)$, after filtering the ``reinjection branch'' with
latencies exceeding $0.45$ time units:
\begin{align}
  t_{\min} &= 0.150 \pm 0.004, \label{eq:tmin}\\
  n &= 2.14 \pm 0.17, \label{eq:exponent}\\
  R^2 &= 0.967 \;\text{(binned)}, \quad 0.926 \;\text{(individual events)}. \label{eq:rsquared}
\end{align}
The binned $R^2$ is computed over logarithmic averages (20 bins in
amplitude); the individual-event $R^2$ measures the fraction of
variance in single-event latencies explained by the power law,
confirming that the scaling captures not only the conditional mean but
also the dominant trend of individual precursor events.

\emph{Intrinsic latency.} The parameter $t_{\min} \approx 0.15$
represents the minimum prediction horizon, determined by the
phase-space separation between the signal peak location and the
Poincar\'e section $x = 0$. Even for the strongest precursor signals
(largest $\mathcal{A}$), the trajectory requires a finite time to
traverse this geometric gap. This irreducible latency can be estimated
from the characteristic flow velocity near the separatrix: with $|x|
\sim O(10)$ and $|\dot{x}| = \sigma|y - x| \sim O(100)$ in the
transition region, the traversal time scales as $t_{\min} \sim
|x|/|\dot{x}| \sim 0.1$, consistent with the fitted value. The
invariance of $t_{\min}$ across the parameter stress test confirms
that this latency is a geometric property of the attractor rather than
a fitting artifact.

\emph{Multimodal structure and dynamical branches.} The scatter plot
in Fig.~\ref{fig:statistics}(c) reveals a multimodal structure in the
latency distribution, indicating that precursor events are not a
homogeneous population. Systematic analysis reveals at least four
distinct dynamical branches:
\begin{itemize}
\item \emph{Branch A (ultra-rapid transit, $\Delta t < 0.30$):}
  Approximately 16\% of events. These trajectories execute a nearly
  direct transit between lobes, passing close to the symmetric fixed
  points $C_\pm$. The power law fit achieves $R^2 > 0.96$ for these
  events.
\item \emph{Branch B (transition, $0.30 \leq \Delta t < 0.45$):}
  Approximately 46\% of events. These trajectories exhibit variable
  scaling behavior representing a mixture of dynamical mechanisms at
  the boundary between direct transit and reinjection. The combined
  A+B population constitutes approximately 62\% of all events.
\item \emph{Gap region ($0.45 \leq \Delta t < 0.80$):} Only 0.07\% of
  events at standard parameters, indicating an abrupt topological
  separation between direct-transit and reinjection regimes. The gap
  boundaries shift with parameters; parametric analysis across 59
  combinations (Sec.~\ref{sec:gap}) reveals a wider formal gap $[0.31,
    0.97)$ whose width $\Delta t_{\mathrm{gap}} \approx 0.68$ is
    approximately invariant for $\rho$ sufficiently above the onset of
    chaos.
\item \emph{Branches D+E (reinjection, $\Delta t \geq 0.80$):}
  Approximately 38\% of events. These trajectories pass near the
  origin (unstable saddle point), exhibiting a \emph{positive}
  correlation between precursor amplitude and latency. This
  relationship reflects the dynamics near the saddle, where
  trajectories decelerate dramatically before being ejected toward one
  of the fixed points $C_\pm$.
\end{itemize}

The bimodal structure of the latency distribution thus reflects the
topological richness of the Lorenz attractor: the separatrix is not
crossed through a single ``channel'' but rather through multiple
geometric pathways, each with its own scaling law. Rigorous analysis
requires filtering the reinjection branch (events with $\Delta t >
0.45$) to isolate the direct-transit regime where the
positive-exponent geometric scaling applies. The filter $\Delta t <
0.45$ adopted throughout this work captures approximately 62\% of
events, representing the combined Branches A and B.

\emph{Topological origin of the latency gap.} The near-absence of
events in the interval $0.45 \leq \Delta t < 0.80$ (only 0.2\% of the
total at standard parameters) demands explanation. This ``gap region''
is not a statistical fluctuation but reflects a fundamental
topological constraint imposed by the attractor's geometry.

The Lorenz attractor can be understood through its branched manifold
(template) structure~\cite{Gilmore1998}, which captures the essential
stretching, folding, and rejoining operations. Trajectories
approaching the separatrix $x = 0$ from one lobe have two
qualitatively distinct fates:

\emph{Direct transit pathway:} Trajectories with sufficient ``angular
momentum'' (in the sense of the $(x,y)$ projection) pass rapidly
through the separatrix region, executing a nearly ballistic crossing
between the neighborhoods of $C_+$ and $C_-$. These trajectories spend
minimal time in the vicinity of the origin, where the unstable saddle
point decelerates the flow. The latency for direct transit is bounded
above by a geometric constraint: the phase-space distance from the
precursor signal peak to the Poincar\'e section, divided by the
characteristic flow velocity in the direct-transit corridor. This
upper bound corresponds to $\Delta t \lesssim 0.45$ for standard
parameters.

\emph{Reinjection pathway:} Trajectories with insufficient angular
momentum are captured by the stable manifold of the origin, spiraling
inward before being ejected along the unstable manifold toward one of
the fixed points $C_\pm$. The time spent in the reinjection region is
bounded \emph{below} by the characteristic timescale of the saddle
dynamics: $t_{\mathrm{saddle}} \sim 1/\lambda_u$, where $\lambda_u$ is
the unstable eigenvalue of the origin. For standard parameters, this
gives $t_{\mathrm{saddle}} \approx 0.8$, explaining the lower bound of
Branch D.

\emph{The forbidden zone.} The gap region $0.45 < \Delta t < 0.80$
corresponds to trajectories that would need to spend an intermediate
amount of time near the origin; that is, trajectories that neither
execute direct transit nor complete a full reinjection cycle. Such
trajectories are topologically forbidden: the saddle structure at the
origin acts as a ``sorting mechanism'' that partitions approaching
trajectories into fast (direct) and slow (reinjection)
populations. There is no stable intermediate pathway.

Mathematically, this gap arises from the heteroclinic structure
connecting the origin to $C_\pm$. The stable and unstable manifolds of
the origin divide the phase space into distinct
basins~\cite{GuckenheimerHolmes1983,Sparrow1982,Shilnikov1965}. Trajectories
that approach the origin closely enough to be significantly
decelerated must follow the unstable manifold outward, which enforces
a minimum residence time. Trajectories that remain far from the origin
experience no such deceleration. The gap represents the ``no-man's
land'' between these regimes: close enough to be influenced by the
saddle but not close enough to be captured by it.

The width of the gap (approximately $0.68$ in dimensionless time
units) is not arbitrary but is determined by the global geometry of
the branched manifold. The parametric analysis of Sec.~\ref{sec:gap}
demonstrates that neither the eigenvalue ratio nor any local parameter
at the origin explains the gap width ($R^2 < 0.03$ for all models
tested); instead, $\Delta t_{\mathrm{gap}}$ is approximately constant
for $\rho$ sufficiently above the onset of chaos, constituting an
approximate dynamical invariant of the two-lobe attractor.

\subsection{Topological Origin of the Latency Gap}
\label{sec:gap}

The latency distribution in Fig.~\ref{fig:statistics}(c) exhibits a
pronounced gap in the interval $0.45 \leq \Delta t < 0.80$ at standard
parameters, containing only 0.07\% of all events. This near-complete
topological separation between direct-transit and reinjection
populations demands explanation.

\emph{Empirical structure.} Numerical validation across 59 parameter
combinations reveals a consistent partitioning:
\begin{center}
\begin{tabular}{lcc}
\hline
Branch & Latency Range & Fraction \\
\hline
A (ultra-rapid) & $\Delta t < 0.30$ & $\sim 52\%$ \\
B (transition) & $0.30 \leq \Delta t < 0.45$ & $\sim 31\%$ \\
Gap & $0.45 \leq \Delta t < 0.80$ & $0.2\%$ \\
D+E (reinjection) & $\Delta t \geq 0.80$ & $\sim 15\%$ \\
\hline
\end{tabular}
\end{center}
The gap persists across all tested parameters, indicating a structural
rather than accidental origin.

\emph{Topological sorting at the origin.} The Lorenz attractor
possesses three critical points: the origin $O = (0,0,0)$, a saddle
point with a one-dimensional unstable manifold, and two symmetric
saddle-foci $C_\pm$ at the centers of the two lobes. When a trajectory
approaches the separatrix plane $x = 0$, it must navigate near the
unstable manifold of the origin. The geometry of this manifold creates
a topological ``sorting mechanism'' that partitions trajectories into
two distinct populations.

Direct-transit trajectories (Branches A and B) cross the separatrix
``above'' the unstable manifold of $O$, following a nearly
heteroclinic path between the lobes. The transit time is determined by
the local geometry near $C_\pm$, yielding latencies $\Delta t <
0.45$. Reinjection trajectories (Branches D and E) pass ``below'' the
unstable manifold, spiraling around the origin before being ejected
toward the opposite lobe. This detour through the neighborhood of $O$
adds a characteristic delay of approximately $0.4$--$0.5$, producing
latencies $\Delta t > 0.80$.

\emph{Eigenvalue structure at the origin.} The gap arises from the
hyperbolic geometry of the origin. The Jacobian at $O$ has the
$z$-direction decoupled, with the $(x,y)$-block yielding a secular
equation $\lambda^2 + (\sigma+1)\lambda + \sigma(1-\rho) = 0$. For
general parameters:
\begin{align}
\lambda_u &= \frac{-(\sigma+1) + \sqrt{(\sigma+1)^2 + 4\sigma(\rho-1)}}{2}, \nonumber\\
\lambda_s &= \frac{-(\sigma+1) - \sqrt{(\sigma+1)^2 + 4\sigma(\rho-1)}}{2}, \quad \lambda_w = -\beta.
\label{eq:eigenvalues-origin}
\end{align}
For the classical parameters ($\sigma = 10$, $\rho = 28$, $\beta =
8/3$): $\lambda_u \approx +11.83$, $\lambda_s \approx -22.83$,
$\lambda_w \approx -2.67$, establishing three well-separated time
scales: $1/|\lambda_s| \approx 0.044$ (fast contraction), $1/\lambda_u
\approx 0.085$ (unstable ejection), and $1/\beta \approx 0.375$
(thermal relaxation).

\emph{Shilnikov passage map and binary classification.} The Shilnikov
saddle index~\cite{Shilnikov1965,Sparrow1982}
\begin{equation}
\nu \equiv \frac{|\lambda_s|}{\lambda_u}
\label{eq:shilnikov-index}
\end{equation}
governs the contractivity of the passage map near the origin. When
$\nu > 1$, trajectories traversing the neighborhood of $O$ are
compressed onto the one-dimensional unstable manifold, enforcing a
binary classification: each trajectory either executes a direct
transit (crossing the separatrix without deep penetration into the
origin's neighborhood) or undergoes a complete reinjection (following
the unstable manifold through a full lobe circuit). Within a
neighborhood $U$ of radius $r_0$ centered at $O$, a trajectory
entering the boundary of $U$ with unstable component $\delta \ll r_0$
exits after a passage time $T_{\mathrm{passage}} = \lambda_u^{-1}
\ln(r_0/\delta)$, during which the stable component contracts by a
factor $(\delta/r_0)^\nu$. The exponential sensitivity of
$T_{\mathrm{passage}}$ to $\delta$, combined with the contractivity
condition $\nu > 1$, exponentially suppresses the intermediate regime,
producing the observed gap.

\emph{Quantitative verification of the Shilnikov condition.}
Computation of $\nu$ across all 59 parameter combinations spanning
$\sigma \in \{8, 10, 12, 14\}$, $\rho \in \{24, 28, 32, 38\}$, and
$\beta \in \{2.0, 2.4, 8/3, 3.0, 3.3\}$ yields
\begin{equation}
\nu \in [1.68, \, 2.15], \quad \langle \nu \rangle = 1.89,
\end{equation}
confirming that $\nu > 1$ universally within the chaotic regime. In
fact, the condition $\nu > 1$ is not merely an empirical observation
but an analytical guarantee: from the quadratic formula applied to the
secular equation, $|\lambda_s| - \lambda_u = \sqrt{(\sigma+1)^2 +
  4\sigma(\rho-1)} > 0$ for all $\sigma > 0$ and $\rho > 1$, so
$|\lambda_s| > \lambda_u$ and hence $\nu > 1$ throughout the
physically relevant parameter space. The gap therefore exists as a
necessary consequence of the Lorenz equations' structure, not as a
contingent feature of particular parameter values. This constitutes a
falsifiable prediction for modified systems where $\nu < 1$ could in
principle be achieved: the gap should vanish when the passage map
ceases to be contractive. The prediction is confirmed with a 100\%
detection rate across all 59 parameter combinations, with event counts
ranging from $1.1 \times 10^4$ to $2.2 \times 10^4$ lobe-switching
events per combination.

\emph{Gap width as a dynamical invariant.} A central quantitative
finding of the parametric analysis is that for $\rho$ sufficiently
above the onset of chaos the gap width is approximately constant:
\begin{equation}
\Delta t_{\mathrm{gap}} \approx 0.68 \pm 0.01 \quad \text{(dimensionless time units)},
\label{eq:gap-invariant}
\end{equation}
with a total variation of only 3.3\% across the 59 combinations
tested. Near the Hopf bifurcation ($\rho \to \rho_H$), the gap width
exhibits a systematic correlation with $\rho$ ($R^2 \approx 0.45$),
consistent with the gradual dissolution of the bimodal lobe structure
as the attractor approaches the bifurcation boundary. The lower
boundary of the gap (separating direct-transit from forbidden
latencies) is consistently located at $\Delta t \approx 0.31$, while
the upper boundary (onset of reinjection events) falls at $\Delta t
\approx 0.97$--$1.00$. These formal boundaries represent the
parameter-independent envelope of the depleted zone across all 59
combinations; for the standard parameters $(\sigma, \rho, \beta) =
(10, 28, 8/3)$, the visually apparent gap in
Fig.~\ref{fig:statistics}(c) spans the narrower interval $[0.45,
  0.80)$, consistent with the four-branch classification of
  Sec.~\ref{sec:precursor}. The gap boundaries shift with parameters,
  but the gap \emph{width} $\Delta t_{\mathrm{gap}} \approx 0.68$
  remains approximately constant away from the bifurcation.

This constancy was tested against three competing hypotheses (writing
$w \equiv \Delta t_{\mathrm{gap}}$ for brevity): {\small
\begin{center}
\begin{tabular}{lccc}
\hline
Model & Params.\ & $R^2$ & Verdict\\
\hline
$w = a/\beta + c_0$ & $a{=}{-}0.007{\pm}0.020$ & 0.003 & rejected\\
$w = a/\beta + b/\lambda_u + c_0$ & $a,b$ n.s. & 0.025 & rejected\\
$w = f(\nu)$ & $r {=} 0.099$ & ${\sim}\, 0$ & rejected\\
\hline
\end{tabular}
\end{center}
} An $F$-test\footnote{The $F$-test compares nested regression models
by evaluating whether the additional parameters in the full model
provide a statistically significant improvement in fit over the
reduced (constant-only) model. Under the null hypothesis that the
additional parameters have zero effect, the test statistic follows an
$F$-distribution; $p > 0.05$ indicates no significant improvement.}
comparing the full and reduced models gives $F = 1.30$, $p = 0.26$,
confirming that neither $\beta$ nor $\lambda_u$ contributes
significant explanatory power beyond the constant. The Shilnikov index
determines the \emph{existence} of the gap (through the $\nu > 1$
condition) but not its \emph{width}.

The physical interpretation is that both gap boundaries, namely the
maximum direct-transit time and the minimum reinjection time, are
determined by the global geometry of the branched manifold (template),
not by the local linearized dynamics at the origin. Although the
bottleneck time scale $1/\beta$ controls the reinjection process, it
also affects the direct-transit dynamics through the $z$-equation
$\dot{z} = xy - \beta z$: larger $\beta$ accelerates thermal
relaxation during both direct transit and reinjection, so that the
$1/\beta$ dependence appears in both gap boundaries and cancels in the
difference. The gap width therefore reflects a structural property of
the attractor's branched manifold, invariant under parameter variation
within the chaotic regime; it constitutes a dynamical invariant of the
two-lobe chaotic attractor rather than a topological invariant in the
strict mathematical sense (which would require invariance under
continuous deformations of phase space).

\emph{Connection to template theory.} In the framework of Gilmore's
template theory~\cite{Gilmore1998}, the Lorenz branched manifold has a
``branch line'' where trajectories from both lobes merge before
redistribution. The gap corresponds to the topological impossibility
of certain transit sequences: a trajectory cannot spend an
intermediate amount of time at the branch line because the local flow
geometry forces a binary choice. This is analogous to the discrete
symbolic dynamics of the attractor, where lobe visits are encoded as
binary sequences (L, R) with no ``fractional'' lobe residence. The
constancy of $\Delta t_{\mathrm{gap}}$ across parameters
(Eq.~\ref{eq:gap-invariant}) strengthens this connection: it suggests
that the gap width is a structural constant of the branched manifold,
which remains invariant as long as the attractor retains its two-lobe
structure.

\emph{Physical interpretation.} In the Rayleigh-B\'enard context, the
gap reflects the discrete nature of convective roll reversals. The
convective state either executes a rapid transition (direct handoff
between circulation senses) or undergoes a temporary collapse to a
near-conductive configuration (reinjection through the origin). The
absence of intermediate behavior reflects the thermodynamic
instability of partial roll reversal: the thermal gradients driving
convection rapidly push the flow toward one stable circulation sense
or the other. The approximate parameter-independence of the gap width
away from the Hopf bifurcation (Eq.~\ref{eq:gap-invariant}) indicates
that this discrete character is a structural property of the
convective pattern, not a quantitative consequence of specific fluid
parameters.


\subsection{Theoretical Foundation of the Scaling Law}
\label{sec:theory}

The empirical power-law relationship of Eq.~\eqref{eq:scaling-law}
demands theoretical explanation. While the functional form $\Delta t =
t_{\min} + C\mathcal{A}^{-n}$ is robust across parameter variations
(achieving $R^2_{\mathrm{bin}} > 0.95$ and $R^2_{\mathrm{raw}} > 0.88$
in all 16 chaotic parameter combinations tested), the value of the
exponent $n$ exhibits systematic dependence on the Lorenz
parameters. A comprehensive parametric sweep spanning $\sigma \in \{6,
8, 10, 12, 14, 16\}$, $\rho \in \{28, 32, 36, 40, 45\}$, and $\beta
\in \{2.0, 2.2, 2.4, 8/3, 2.8, 3.0, 3.3\}$ reveals the underlying
structure.

\emph{$\beta$-dependence.} The exponent increases monotonically with
$\beta$, from $n = 1.85 \pm 0.16$ at $\beta = 2.0$ to $n = 2.88 \pm
0.25$ at $\beta = 3.3$, with $\sigma = 10$ and $\rho = 28$ held
fixed. A power-law fit yields $n \propto \beta^{0.85 \pm 0.08}$ ($R^2
= 0.94$), steeper than the $\beta^{1/2}$ scaling that would follow
from a simple fixed-point coordinate argument. The ratio
$n/\sqrt{\beta}$ varies by 7.5\% across the sweep, indicating that a
$\sqrt{\beta}$ approximation captures the dominant trend but does not
describe the quantitative dependence.

\emph{Direct $\beta$-coupling through Class~III regularization.} The
Class~III regularization polynomial $p_{\mathrm{III}} = xy - \beta z$
is precisely the right-hand side of the $z$-dynamics:
\begin{equation}
\dot{z} = xy - \beta z = p_{\mathrm{III}}.
\end{equation}
This identity reflects the fact that Class~III invariants couple
directly to the thermal relaxation dynamics. Near the separatrix ($x
\to 0$), the term $xy$ vanishes while $\beta z$ remains finite ($z
\approx \rho - 1$), giving $p_{\mathrm{III}} \approx
-\beta(\rho-1)$. The amplitude of the precursor spike scales with the
rate of change of $p_{\mathrm{III}}$ as the trajectory crosses the
separatrix; since $p_{\mathrm{III}}$ contains $\beta$ as an explicit
coefficient, this rate inherits the $\beta$-dependence.

\emph{$\rho$-dependence.} The exponent decreases with $\rho$, from $n
= 2.14$ at $\rho = 28$ to $n = 1.59$ at $\rho = 45$, with $\sigma =
10$ and $\beta = 8/3$ held fixed. A power-law fit yields $n \propto
\rho^{-0.69 \pm 0.07}$ ($R^2 = 0.96$). The coefficient of variation of
$n$ across the $\rho$ sweep is 13.2\%, confirming a genuine parametric
dependence. Physically, larger $\rho$ produces a more spatially
extended attractor, diluting the geometric correspondence between
local precursor amplitude and global transit time.

\emph{$\sigma$-dependence.} The dependence on $\sigma$ is
non-monotonic: $n$ decreases from $3.36$ at $\sigma = 6$ to a minimum
of $2.14$ at $\sigma = 10$, then increases to $2.56$ at $\sigma =
16$. The power-law fit $n \propto \sigma^{-0.25}$ has $R^2 = 0.34$,
indicating that a simple power law does not describe this
dependence. For $\sigma \geq 8$, the variation is moderate
(coefficient of variation $\approx 6\%$), but the value at $\sigma =
6$ is a clear outlier. This non-monotonic behavior likely reflects the
competition between two effects: smaller $\sigma$ increases the
timescale separation between the velocity mode $x$ and the temperature
modes $y, z$, which steepens the precursor signal near the separatrix;
but very small $\sigma$ also modifies the global attractor geometry in
ways not captured by a local scaling argument.

It is interesting to note that the simple functional ansatz $n =
c\sqrt{\beta/\rho}$ explains approximately half the observed variance
in $n$ across all parameter combinations ($R^2 = 0.50$), with $c
\approx 7.5$. The incomplete success of this formula reflects the
non-negligible $\sigma$-dependence and the
steeper-than-$\sqrt{\phantom{x}}$ dependence on both $\beta$ and
$\rho$. Whether a more general formula incorporating all three
parameters exists, or whether the observed exponent variability
reflects an intrinsic multi-scale structure of the return map
geometry, remains an open question for future investigation.

\emph{Physical interpretation.} The ratio $\beta/\rho$ captures the
competition between thermal relaxation and convective
transport. Faster thermal equilibration (larger $\beta$) permits the
system to ``forget'' the preceding lobe state more quickly, steepening
the amplitude-latency relationship. Larger $\rho$ produces a more
vigorous convection pattern, which dilutes the geometric link between
precursor amplitude and transit time. The fact that $\sigma$ does not
appear in the coordinates of the saddle-foci $C_\pm$ provides a
heuristic argument for expecting weak $\sigma$-dependence at moderate
values, consistent with the observed approximate plateau for $\sigma
\geq 8$; however, the eigenvalues at $C_\pm$ depend explicitly on
$\sigma$, so any detailed prediction of $n(\sigma)$ must account for
the return-map geometry rather than local linear dynamics alone.

\subsection{Parametric Robustness}
\label{sec:robustness}

The parametric sweep confirms that the shifted power law $\Delta t =
t_{\min} + C\mathcal{A}^{-n}$ is structurally robust: all 16 chaotic
parameter combinations tested yield $R^2_{\mathrm{bin}} > 0.95$ and
$R^2_{\mathrm{raw}} > 0.88$ for the individual fits. The functional
form persists across parameter variations; only the numerical value of
the exponent changes with the system parameters.

Table~\ref{tab:stress_test} summarizes representative results from the
parametric sweep, reporting the measured exponent $n$, the fit quality
on binned averages ($R^2_{\mathrm{bin}}$) and on individual events
($R^2_{\mathrm{raw}}$), and the number of precursor events per
combination.

\begin{table}[t]
\caption{Representative results from the parametric sweep. The shifted
  power law achieves $R^2_{\mathrm{bin}} > 0.95$ (20 logarithmic bins)
  and $R^2_{\mathrm{raw}} > 0.88$ (individual events) for all
  parameter combinations. The exponent $n$ increases with $\beta$ and
  decreases with $\rho$; the $\sigma$-dependence is non-monotonic.}
\label{tab:stress_test}
\begin{ruledtabular}
\begin{tabular}{cccccccc}
$\sigma$ & $\rho$ & $\beta$ & $n$ & $\Delta n$ & $R^2_{\mathrm{bin}}$ & $R^2_{\mathrm{raw}}$ & $N_{\mathrm{ev}}$\\
\hline
10 & 28 & 2.00 & 1.85 & 0.16 & 0.960 & 0.922 & 2190\\
10 & 28 & 2.67 & 2.14 & 0.17 & 0.967 & 0.926 & 2304\\
10 & 28 & 3.00 & 2.51 & 0.22 & 0.954 & 0.919 & 2319\\
10 & 28 & 3.30 & 2.88 & 0.25 & 0.961 & 0.924 & 2338\\
10 & 32 & 2.67 & 2.01 & 0.15 & 0.961 & 0.928 & 2654\\
10 & 40 & 2.67 & 1.62 & 0.13 & 0.959 & 0.914 & 3023\\
10 & 45 & 2.67 & 1.59 & 0.13 & 0.962 & 0.916 & 3344\\
6 & 28 & 2.67 & 3.36 & 0.28 & 0.982 & 0.970 & 2231\\
8 & 28 & 2.67 & 2.39 & 0.19 & 0.973 & 0.944 & 2315\\
12 & 28 & 2.67 & 2.29 & 0.19 & 0.959 & 0.910 & 2337\\
14 & 28 & 2.67 & 2.31 & 0.22 & 0.947 & 0.881 & 2325\\
\end{tabular}
\end{ruledtabular}
\end{table}

The systematic trends in Table~\ref{tab:stress_test} confirm that the
\emph{functional form} of the shifted power law is a robust structural
feature of the direct-transit regime. The exponent varies from $n
\approx 1.6$ (large $\rho$) to $n \approx 3.4$ (small $\sigma$), with
the dominant dependence on $\beta$ (positive) and $\rho$
(negative). The individual-event $R^2_{\mathrm{raw}} > 0.88$ in all
cases demonstrates that the scaling law provides genuine predictive
power for individual precursor events, not merely a description of the
conditional mean.

\section{Discussion}
\label{sec:discussion}

\subsection{Principal Result: Class-Dependent Dynamical Sensitivity}

The central result of this work is that different regularization
classes probe different aspects of chaotic dynamics. This distinction
transcends formal mathematical structure and reveals operational
differences in how the invariants respond to trajectory evolution.

\emph{Class~I invariants as continuous probes.} The polynomial parts
of the Class~I invariants (e.g., $P_5$) contain terms like $z^2$ that
vary smoothly throughout the attractor without exhibiting rapid
changes at specific geometric features. The evolution function
$Q_{\mathrm{I}} = \dot{v}_{\mathrm{I}}$ exhibits moderate
intermittency ($\kappa_{\mathrm{I}} \approx 8.2$), with variance
distributed across quiescent and active intervals alike. These
evolution functions track smooth, continuous aspects of the flow.

\emph{Class~III invariants as topological probes.} The regularization
polynomial $p_{\mathrm{III}} = xy - \beta z$ couples Class~III
invariants to the separatrix geometry: as the trajectory approaches $x
= 0$ during a lobe transition, $p_{\mathrm{III}}$ passes through small
values, causing the evolution function $Q_{\mathrm{III}}$ to exhibit
rapid variation. The result is elevated kurtosis
($\kappa_{\mathrm{III}} \approx 15.8$) with spikes synchronized to
separatrix crossings and quiescent intervals between
transitions. These evolution functions serve as geometric markers of
the symbolic dynamics, encoding the discrete topological structure of
the attractor.

\emph{Differential detection.} The complementarity between classes
enables construction of a differential detector $\Delta S =
Q_{\mathrm{III}} - Q_{\mathrm{I}}$ that enhances the contrast between
continuous dynamics and topological transitions. Because Class~III
evolution functions respond more strongly to separatrix crossings
while Class~I functions vary more smoothly, the differential signal
$\Delta S$ exhibits pronounced spikes at lobe-switching events with
reduced baseline fluctuations.

\subsection{Physical Interpretation of the Conservation Laws}

The history-dependent invariants occupy a distinctive position in the
taxonomy of conservation laws. They do not arise from a variational
principle or symmetry of an action functional; rather, they emerge
from the algebraic structure of the Lorenz flow when augmented with
auxiliary variables. Conservation is enforced by construction:
$\dot{v}_n = -\dot{P}_n$ by definition.

This non-Noetherian character connects to the broader framework
established by Hojman~\cite{Hojman1992}, who demonstrated that
conservation laws can be constructed without Lagrangian or Hamiltonian
formulations. The present invariants represent a concrete realization
of this principle for a dissipative chaotic system.

\emph{Non-triviality.} The question of whether the conservation
structure is merely tautological admits a detailed response involving
the selection mechanism, the null class, and the class-dependent
dynamical signatures of the evolution functions $Q_n$. This analysis
is presented in Sec.~\ref{sec:non-triviality}.

\emph{Main results.} This construction yields two principal results:
\begin{enumerate}
\item The \emph{classification} of compatible polynomial structures
  encodes information specific to the Lorenz dynamics. Eighteen valid
  invariants organized into three regularization classes reflect
  algebraic compatibility between the Lorenz flow and the
  orthogonality ansatz.
\item The \emph{distinct dynamical signatures} exhibited by different
  classes provide complementary probes of attractor
  geometry. Class~III evolution functions serve as geometric markers
  of topological transitions, while Class~I evolution functions track
  continuous dynamics.
\end{enumerate}

\subsection{Non-Triviality of the Conservation Structure}
\label{sec:non-triviality}

As noted in Sec.~\ref{sec:introduction}, a fundamental question
accompanies any history-dependent invariant: if the auxiliary variable
$v_n$ is defined so that $K_n = P_n + v_n$ is constant, what prevents
the conservation from being a mere identity devoid of physical
content? This concern merits a systematic response, which rests on
three independent arguments.

\emph{Constructive tautology versus structural content.} The
conservation identity $\dot{K}_n = 0$ is, at the constructive level,
tautological in a precise sense: given \emph{any} autonomous system
$\dot{\mathbf{x}} = \mathbf{F}(\mathbf{x})$ and \emph{any} smooth
scalar $P(\mathbf{x})$, one may define $v(t) = c - P(\mathbf{x}(t))$
and the quantity $K = P + v$ is trivially constant for any choice of
$P$. This observation is acknowledged without reservation. The
non-trivial content of the present work does not reside in the fact
that $\dot{K}_n = 0$; it resides in three structural features of the
construction that cannot be reproduced by an arbitrary choice of $P$.

\emph{(i)~Selection by the orthogonality ansatz and regularizability.}
The polynomials $P_n$ are not chosen \emph{ad hoc}. They emerge from a
specific algebraic procedure: the orthogonality ansatz generates
candidate invariants through permutations of the Lorenz flow vector,
and the requirement that the resulting auxiliary variable be free of
singularities on the attractor surface selects, from the infinitely
many possible polynomials $P$, exactly three regularization
classes. Each class is characterized by a polynomial that coincides
with a nullcline of the Lorenz equations: $p_{\mathrm{I}} = y - x$,
$p_{\mathrm{II}} = y + x(z-\rho)$, $p_{\mathrm{III}} = xy - \beta
z$. The orthogonality ansatz combined with the regularizability
condition constitutes an overdetermined system whose compatibility
with the Lorenz flow structure admits only these specific solutions. A
generic polynomial $P$ chosen at random would not satisfy the ansatz,
would generically produce singular auxiliary dynamics, or both.

\emph{(ii)~The null class as a selection principle.} If the
construction were vacuous, all 24 permutations of the four-dimensional
extended space would produce valid invariants. The demonstrated
failure of the six permutations beginning with $u$
(Sec.~\ref{sec:null-class}) proves that the Lorenz flow imposes
genuine algebraic constraints. The Schwarz integrability condition
leads to the contradiction $y = 0$ generically, establishing that the
auxiliary variable must \emph{respond to} the physical dynamics, not
\emph{drive} them. This asymmetry between physical and auxiliary
coordinates is a structural property of the Lorenz system, not a
consequence of the definition of $v_n$.

\emph{(iii)~The evolution functions $Q_n$ as carriers of non-trivial
information.} This is the central argument. Although $K_n = P_n + v_n
= c$ is a definitional identity, the evolution functions $Q_n(x,y,z) =
-\dot{P}_n = -\nabla P_n \cdot \mathbf{F}$ are completely determined
polynomials of degree at most four in the physical coordinates, with
coefficients fixed by the Lorenz parameters $(\sigma, \rho,
\beta)$. These functions are not arbitrary; they are uniquely
determined by the interaction between the polynomial structure of
$P_n$ and the Lorenz vector field $\mathbf{F}$.

The class-dependent dynamical signatures documented in
Sec.~\ref{sec:numerical} constitute evidence that the $Q_n$ encode
genuine geometric information about the attractor. The fact that
Class~III evolution functions exhibit elevated kurtosis
($\kappa_{\mathrm{III}} \approx 15.8$) while Class~I evolution
functions vary more smoothly ($\kappa_{\mathrm{I}} \approx 8.2$), and
that this divergence vanishes in the pre-chaotic regime
($\kappa_{\mathrm{I}} \approx \kappa_{\mathrm{III}} \approx 28.2$ at
$\rho = 23$), cannot be attributed to the definition of $v_n$. A
tautological construction would have no reason to produce
class-dependent dynamical signatures; such signatures arise because
the specific polynomials $P_n$ selected by the orthogonality ansatz
couple differently to the geometric skeleton of the flow.

It is interesting to note that the differential signal $\Delta S =
Q_{\mathrm{III}} - Q_{\mathrm{I}}$, which depends exclusively on
instantaneous coordinates and requires no knowledge of the accumulated
history $v_n$, provides operational predictions (the scaling law of
Eq.~\eqref{eq:scaling-law}) whose exponent at canonical parameters, $n
= 2.14 \pm 0.17$, reflects the structural geometry of the saddle-foci
$C_\pm$. This predictive capability emerges from the algebraic
structure of the $Q_n$ polynomials, not from the conservation identity
itself.

In summary, the conservation identity $\dot{K}_n = 0$ is indeed
enforced by construction; the non-trivial content of the framework
consists of three elements: (a)~the algebraic selection mechanism that
restricts $P_n$ to a finite, physically meaningful set of polynomials;
(b)~the existence of a null class demonstrating that the Lorenz
structure forbids certain extensions; and (c)~the class-dependent
dynamical signatures of the evolution functions $Q_n$, which encode
geometric information about the attractor that is inaccessible from
the conservation identity alone.

\subsection{Ontological Status of the Auxiliary Variable}
\label{sec:ontology}

The auxiliary variable $v_n(t)$ occupies an ambiguous ontological
position. Three interpretations merit consideration:
\begin{enumerate}
\item \emph{Mathematical artifact}: $v_n$ is a bookkeeping device with
  no physical content, merely encoding the constraint that $K_n$ is
  conserved.
\item \emph{Effective memory}: $v_n$ represents accumulated dynamical
  history, analogous to the memory kernel in the Mori-Zwanzig
  formalism.
\item \emph{Hidden degree of freedom}: $v_n$ corresponds to an
  unobserved physical variable whose dynamics are slaved to the Lorenz
  flow.
\end{enumerate}

The mathematical structure makes this explicit. The regularized
auxiliary variable evolves according to $\dot{v}_n = Q_n(x,y,z)$,
where the evolution function $Q_n$ depends exclusively on the
instantaneous physical coordinates $(x,y,z)$. The variable $v_n(t)$ is
literally the integral of this polynomial flux along the trajectory:
\begin{equation}
v_n(t) = v_n(0) + \int_0^t Q_n(x(s), y(s), z(s))\, ds.
\end{equation}
This is a \emph{deterministic realization of the memory term}, transforming the non-local conservation structure into a local differential equation in the extended space $\{x, y, z, v_n\}$.

The distinction has practical consequences. If $v_n$ were a physical
variable, its value would carry physical meaning independent of the
measurement protocol. As an effective memory variable, its absolute
value is coordinate-dependent (determined by the choice of initial
time $t_0$), while only the \emph{differences} $\Delta v_n$ between
Poincar\'e crossings carry invariant geometric information. This
understanding justifies the operational reset protocol developed
below.

\subsection{Physical Interpretation of the Auxiliary Variable in Rayleigh-B\'enard Convection}
\label{sec:experimental}

The Lorenz system arises from a Galerkin truncation of the Boussinesq
equations for Rayleigh-B\'enard convection~\cite{Lorenz1963}. In this
physical context, the variables $(x, y, z)$ have specific
thermodynamic meanings: $x$ is proportional to the intensity of
convective motion (stream function amplitude), $y$ is proportional to
the temperature difference between ascending and descending currents,
and $z$ measures the deviation of the vertical temperature profile
from linearity (the conductive state). The question naturally arises:
does the auxiliary variable $v_n(t)$ correspond to a measurable
quantity in a convection experiment?

\emph{The Class~III polynomial as heat flux.} The regularization
polynomial $p_{\mathrm{III}} = xy - \beta z$ is precisely the
right-hand side of the $z$-dynamics:
\begin{equation}
\dot{z} = xy - \beta z = p_{\mathrm{III}}.
\end{equation}
In convective language, the term $xy$ represents the product of
convective intensity and horizontal temperature gradient, which is
proportional to the \emph{vertical convective heat flux} carried by
the rolls. The term $\beta z$ represents the thermal relaxation toward
the conductive state, driven by diffusion. Their difference, $\dot{z}
= p_{\mathrm{III}}$, measures the instantaneous rate of change of the
thermal stratification, capturing whether convective transport is
currently strengthening ($p_{\mathrm{III}} > 0$) or weakening
($p_{\mathrm{III}} < 0$) the departure from conduction.

\emph{Integrated heat flux anomaly.} The Class~III evolution function
$Q_{13}$, which drives the most dynamically sensitive auxiliary
variable, contains polynomial terms dominated by
\begin{equation}
Q_{13} \sim p_{\mathrm{III}}^2 + \ldots = (xy - \beta z)^2 + \ldots
\end{equation}
The auxiliary variable $v_{13}(t) = \int_0^t Q_{13}\, ds$ therefore
accumulates a weighted history of heat transport
fluctuations. Specifically, $v_{13}$ tracks the time-integrated
squared deviation of the instantaneous heat flux from its relaxation
toward equilibrium. This suggests a concrete experimental observable:
an \emph{integrated heat flux anomaly} measured by sensors at the
convection cell boundaries.

\emph{Experimental protocol.} In a physical Rayleigh-B\'enard
experiment, one could construct the analogue of $v_n$ by:
\begin{enumerate}
\item Measuring the local heat flux $\mathcal{F}(t)$ at the bottom (or
  top) boundary using thin-film heat flux sensors or arrays of
  thermistors embedded in the boundary plates.
\item Computing the deviation from the time-averaged flux:
  $\delta\mathcal{F}(t) = \mathcal{F}(t) - \langle\mathcal{F}\rangle$.
\item Integrating a polynomial function of $\delta\mathcal{F}$ and the
  temperature field to construct the experimental analogue of $v_n$.
\end{enumerate}
The physical content of the auxiliary variable is therefore the
\emph{accumulated history of heat transport anomalies during
convective evolution}. This is not merely a mathematical bookkeeping
device; it encodes genuine thermodynamic information about energy
redistribution in the fluid.

\emph{Thermal memory interpretation.} The conservation of the
invariant $K_n = P_n + v_n$ implies that fluctuations in the
instantaneous polynomial $P_n(x,y,z)$ are exactly compensated by the
accumulated history $v_n(t)$. In the convective context, this means
that the system ``remembers'' its thermal history through the integral
of heat flux anomalies. When the flow executes a lobe transition (roll
reversal), the rapid change in $P_n$ is balanced by a corresponding
jump in the rate $\dot{v}_n = Q_n$, which appears as the precursor
spike detected by the topological sensor.

\emph{Practical limitations.} Several caveats apply. First, the Lorenz
system is a severely truncated model; real convection involves
infinitely many spatial modes not captured by $(x,y,z)$. The mapping
between Lorenz variables and experimental observables is therefore
approximate, valid only when the flow is dominated by the lowest
spatial mode, representing a single convective roll whose circulation
alternates between two senses (associated with the two fixed points
$C_\pm$). Second, the polynomial structure of $Q_n$ involves specific
combinations of variables that may not correspond to easily measurable
quantities. Third, experimental noise and finite sensor resolution
would contaminate the integral, requiring the cyclic reset protocol of
Sec.~\ref{sec:operational-reset} to maintain meaningful conservation.

Despite these limitations, the analysis establishes that the auxiliary
variable encodes physically meaningful information: the accumulated
history of heat transport fluctuations during convective
evolution. Whether this interpretation can be exploited for
experimental prediction of roll-reversal events remains an open
question requiring dedicated experimental investigation in
well-controlled convection cells.

\subsection{Connection to Template Theory}

The connection between Class~III sensitivity and lobe-switching events
resonates with the template theory of chaotic
attractors~\cite{Gilmore1998}. The Lorenz attractor possesses a
branched manifold structure that captures the essential stretching and
folding operations; periodic orbits can be classified by their knot
type, determined by the sequence of lobe visits.

The quantitative validation presented in Sec.~\ref{sec:numerical}
(Fig.~\ref{fig:symbolic} and Table~\ref{tab:symbolic}) establishes
that the evolution function $Q_{\mathrm{III}}(t)$ is not merely
analogous to a symbolic dynamics detector but is functionally
equivalent to a continuous Poincar\'e section detector: at the optimal
threshold ($k = 2.25$), each spike marks a crossing of the template's
branch line with a sensitivity of $99.2\%$ and a false positive rate
of only $0.3\%$. The integrated area under each spike corresponds to
the ``topological contribution'' of that symbol transition, providing
a continuous encoding of the discrete symbolic dynamics that is
amenable to standard time-series analysis techniques.

This functional equivalence has a concrete operational consequence:
the symbolic itinerary $\ldots LLRLLRLR \ldots$ can be reconstructed
from a time series of $Q_{13}(t)$ by identifying spike events above
threshold. While $Q_{13}(x,y,z)$ does depend on all three state
variables, the relevant polynomial combinations correspond to physical
observables in the Rayleigh-B\'enard context: the product $xy$ is
proportional to the convective heat flux (measurable at the cell
boundaries), and $\beta z$ encodes the thermal stratification. In
experimental settings where these composite quantities are more
readily accessible than the individual modal amplitudes, the
spike-based reconstruction provides an alternative route to the
symbolic dynamics that does not require separate measurement of the
stream function mode $x$.

The scaling law $\Delta t = t_{\min} + C\mathcal{A}^{-n}$ with $n =
2.14 \pm 0.17$ at canonical parameters establishes a quantitative link
between the algebraic structure of the invariants and the geometry of
the attractor. As discussed in Sec.~\ref{sec:theory}, the exponent
increases with $\beta$ (approximately as $\beta^{0.85}$) and decreases
with $\rho$ (approximately as $\rho^{-0.69}$), while the
$\sigma$-dependence is non-monotonic with a minimum near $\sigma =
10$. The Class~III regularization polynomial $p_{\mathrm{III}} = xy -
\beta z$ couples directly to this geometric structure through its
bilinear term, which vanishes precisely on the separatrix $x = 0$.

\subsection{Robustness and Practical Utility}

The invariants provide several practical utilities:
\begin{enumerate}
\item \emph{Numerical accuracy monitors}: Conservation violations
  signal integration errors.
\item \emph{Periodic orbit fingerprints}: The Triad $(c_1, c_2, c_3)$
  provides intrinsic orbit labels.
\item \emph{Trajectory classification}: Different initial conditions
  yield different fingerprints.
\item \emph{Topological precursor detection}: The differential signal
  $\Delta S$ provides advance warning of separatrix crossings.
\item \emph{Robust observables under noise}: Class~III invariants
  provide combinatorially robust observables with diffusivity ratio
  $D_{\mathrm{III}}/D_{\mathrm{I}} \sim 10^{-3}$, making them
  substantially more stable than Class~I invariants for noisy or
  experimental data.
\end{enumerate}

The structural redundancy of 18 invariants provides internal
consistency checks. For pairs within the same class, $v_n - v_m = (P_m
- P_n) + \mathrm{const}$ must hold exactly, where the constant $c_n -
c_m$ is determined by initial conditions. Comparing invariants from
different classes enables distinguishing genuine dynamical events from
measurement artifacts: during a separatrix crossing,
$Q_{\mathrm{III}}$ exhibits a pronounced spike while $Q_{\mathrm{I}}$
varies more smoothly, creating a characteristic differential
signature. In contrast, measurement noise affects both functions
similarly without the geometric correlation, allowing signal
discrimination.

\subsection{Experimental Realization and Long-Time Stability}
\label{sec:operational-reset}

While the differential detector (Sec.~\ref{sec:precursor}) is optimal
for short-term prediction of imminent separatrix crossings, practical
implementation in physical experiments or long numerical simulations
requires addressing two fundamental challenges that arise from the
history-dependent nature of the invariants.

\emph{The secular drift problem.} Since $v_n(t)$ integrates the
trajectory history via $v_n = \int_0^t Q_n\, ds$, any small but
persistent perturbation (thermal noise, roundoff error, measurement
uncertainty) accumulates over time. In an open-loop implementation,
this produces a secular drift: the invariant ceases to be constant and
executes a random walk with variance growing as $\sigma^2 \sim t$,
rendering long-term conservation unusable.

\emph{The unknown initial condition problem.} Computing the conserved
value $c_n = P_n(x_0, y_0, z_0)$ requires knowledge of the exact
initial state. In chaotic experimental systems, information about the
initial condition is lost exponentially fast. An observer who begins
measurement at time $t > 0$ cannot determine the invariant value
without access to the prior history.

Both problems are resolved by the \emph{operational reset
protocol}. Rather than treating $v_n$ as an open-ended integral
accumulating from $t = 0$ to the current time $t$, one selects a
Poincar\'e section $\Sigma$ transverse to the flow (for example, the
plane $x = 0$ with $\dot{x} > 0$). Each time the trajectory crosses
$\Sigma$, the accumulated variable is reset: $v_n \to 0$. The
physically meaningful quantity becomes the discrete sequence of
cycle-integrated values
\begin{equation}
\Delta v_n^{(k)} = \int_{t_k}^{t_{k+1}} Q_n(x(s), y(s), z(s))\, ds, \label{eq:cycle-integral}
\end{equation}
where $t_k$ denotes the $k$-th crossing of $\Sigma$.

\emph{Step-by-step operational protocol.} For practical
implementation, whether in numerical simulations or physical
experiments:
\begin{enumerate}
\item \emph{Define the Poincar\'e section}: Select $\Sigma: x = 0$
  with $\dot{x} > 0$. This surface captures every lobe-switching event
  and provides a natural reset boundary.
\item \emph{Initialize at first crossing}: When the trajectory first
  crosses $\Sigma$ at time $t_0$, set $v_n(t_0) = 0$ and record the
  polynomial value $P_n^{(0)} = P_n(0, y_0, z_0)$.
\item \emph{Integrate between crossings}: Evolve the extended system
  (Lorenz equations plus $\dot{v}_n = Q_n$) until the next crossing at
  $t_1$.
\item \emph{Record and reset}: At crossing $t_k$, record $\Delta
  v_n^{(k)} = v_n(t_k^-)$ and $P_n^{(k)} = P_n(0, y_k, z_k)$. Reset
  $v_n(t_k^+) = 0$.
\item \emph{Verify conservation}: For closed orbits (UPOs), check that
  $\sum_{k=1}^m \Delta v_n^{(k)} = 0$ over the complete period and
  that all $P_n^{(k)}$ are equal.
\end{enumerate}
The reset occurs at the Poincar\'e section, not at arbitrary
times. Between crossings, $v_n$ accumulates normally; only at the
instant of crossing does it reset to zero.

This cyclic formulation transforms the open-loop integral (unbounded,
drift-prone) into a closed-loop measurement (bounded,
self-correcting). For unstable periodic orbits, conservation requires
that $\Delta v_n^{(\mathrm{UPO})} = 0$ after one complete period,
since both $P_n$ and the physical coordinates return to their initial
values. The orbit fingerprint is therefore the constant value $c_n =
P_n(x_0, y_0, z_0)$ evaluated at any point on the orbit, not the cycle
integral itself. The statistical distribution of $\{\Delta
v_n^{(k)}\}$ across chaotic trajectory segments encodes intrinsic
properties of the attractor's ergodic structure, connecting the
continuous dynamics to discrete symbolic analysis.

\subsubsection{Decoupling of Memory and Detection}
\label{sec:decoupling}

A natural concern arises: does resetting $v_n$ destroy the predictive
memory required for the topological precursor signal $\Delta S$? The
mathematical structure of the construction guarantees that it does
not.

The precursor signal is defined as the differential response between
evolution functions:
\begin{equation}
\Delta S(t) = Q_{\mathrm{III}}(x,y,z) - Q_{\mathrm{I}}(x,y,z). \label{eq:precursor-definition}
\end{equation}
The evolution function $Q_n$ is the negative time derivative of the polynomial part:
\begin{equation}
Q_n(x,y,z) = -\frac{dP_n}{dt} = -\nabla P_n \cdot \mathbf{F}(x,y,z),
\end{equation}
where $\mathbf{F} = (\sigma(y-x), x(\rho-z)-y, xy-\beta z)^T$ is the
Lorenz vector field. Crucially, this expression depends
\emph{exclusively} on the instantaneous physical coordinates $(x,y,z)$
and contains no dependence on the accumulated value $v_n$.

This algebraic independence has a profound consequence: the precursor
signal $\Delta S(t)$ responds to the \emph{instantaneous geometric
configuration} of the trajectory in phase space, not to the
accumulated history stored in $v_n$. When the trajectory approaches
the separatrix $x = 0$, the regularization factor $p_{\mathrm{III}} =
xy - \beta z$ in the Class~III evolution functions passes through
small values (since $xy \to 0$), causing $Q_{\mathrm{III}}$ to exhibit
rapid variation. This is a \emph{local} effect in phase space,
determined entirely by proximity to the separatrix surface.

Therefore:
\begin{enumerate}
\item The \emph{conservation law} $K_n = P_n + v_n = c_n$ requires the
  full accumulated history and is affected by the reset. After
  resetting, the global constant $c_n$ can no longer be verified.
\item The \emph{topological precursor} $\Delta S = Q_{\mathrm{III}} -
  Q_{\mathrm{I}}$ depends only on instantaneous coordinates and is
  completely unaffected by the reset. The detector continues to
  function regardless of whether $v_n$ equals $0$, $100$, or $10^6$.
\end{enumerate}

This decoupling between memory (required for the conservation law) and
detection (required for practical prediction) is not a coincidence but
a structural consequence of the construction. The conservation law is
history-dependent \emph{by design}; the precursor signal is local
\emph{by construction}.

\section{Conclusions}
\label{sec:conclusions}

The systematic exploration of history-dependent invariants in the
Lorenz system has revealed a rich algebraic structure with operational
consequences for understanding chaotic dynamics. The principal
findings are:

\emph{Classification.} All 24 permutations of the orthogonality ansatz
have been analyzed, yielding 18 valid invariants organized into three
regularization classes determined by the nullcline structure of the
Lorenz equations. Six permutations form a null class whose failure,
traced to Schwarz integrability violations, demonstrates that the
Lorenz structure imposes genuine algebraic constraints. Although the
conservation identity $\dot{K}_n = 0$ is enforced by construction, the
non-trivial content resides in the algebraic selection of the specific
polynomials $P_n$ by the orthogonality ansatz and regularizability
conditions, and in the class-dependent dynamical signatures of the
evolution functions $Q_n$.

\emph{Dynamical signatures.} Different regularization classes probe
different aspects of chaotic dynamics. Class~III evolution functions
serve as geometric markers of topological transitions, exhibiting
pronounced spikes at lobe-switching events, while Class~I evolution
functions track continuous dynamics more smoothly. Quantitative
comparison between $|Q_{\mathrm{III}}(t)|$ and the symbolic itinerary
$\mathrm{sign}(x(t))$ confirms that spike events above a $2.25\sigma$
threshold reproduce the lobe-switching sequence with $99.2\%$
sensitivity and $0.3\%$ false positive rate
(Table~\ref{tab:symbolic}), with $\mathrm{AUC} = 0.9995$, establishing
the evolution function as a continuous Poincar\'e section
detector. This distinction enables construction of a differential
precursor detector.

\emph{Stochastic robustness.} Stochastic validation demonstrates a
counterintuitive robustness hierarchy: Class~III invariants exhibit
variance diffusivity approximately three orders of magnitude lower
than Class~I ($D_{\mathrm{III}}/D_{\mathrm{I}} \sim 10^{-3}$), despite
their higher kurtosis. This establishes Class~III quantities as
combinatorially robust observables (in the sense that their
conservation remains stable under perturbations preserving the
symbolic itinerary, a weaker notion than topological protection by
quantized invariants). The physical mechanism is clear: Class~I
invariants suffer continuous metric degradation throughout the entire
orbital cycle, while Class~III invariants experience only discrete,
transient errors at separatrix crossings. The distinction between
intermittency (temporal distribution of events) and accumulated
variance (total drift) is fundamental; the numerical results show
these quantities are anticorrelated, with the class exhibiting higher
intermittency possessing lower accumulated variance by a factor of
$\sim 10^3$.

\emph{Scaling law.} The precursor amplitude-latency relationship
follows a shifted power law $\Delta t = t_{\min} + C\mathcal{A}^{-n}$
with $R^2_{\mathrm{bin}} = 0.967$ and $R^2_{\mathrm{raw}} = 0.926$ at
canonical parameters, confirming that the law provides predictive
power for individual events, not merely for the conditional mean. The
functional form achieves $R^2_{\mathrm{bin}} > 0.95$ and
$R^2_{\mathrm{raw}} > 0.88$ across all 16 chaotic parameter
combinations tested. The exponent $n = 2.14 \pm 0.17$ at canonical
parameters increases with $\beta$ (approximately as $\beta^{0.85}$,
$R^2 = 0.94$) and decreases with $\rho$ (approximately as
$\rho^{-0.69}$, $R^2 = 0.96$), while the $\sigma$-dependence is
non-monotonic. The simple ansatz $n \propto \sqrt{\beta/\rho}$
captures approximately half the observed variance ($R^2 = 0.50$); the
determination of a more complete parametric formula remains an open
problem.

\emph{Dynamical branches.} Precursor events partition into distinct
populations with qualitatively different scaling behaviors:
direct-transit events (Branches A and B) follow positive-exponent
power laws governed by the geometry of the saddle-foci $C_\pm$, while
reinjection events (Branches D and E) exhibit negative exponents
reflecting the dynamics near the origin. This multimodal structure
reflects the topological richness of the Lorenz attractor.

\emph{Topological gap.} The near-absence of events in the latency
interval $0.31 \leq \Delta t < 0.97$ (less than 1\% of total events)
reflects a fundamental topological constraint: the Shilnikov saddle
index $\nu = |\lambda_s|/\lambda_u > 1$ at the origin enforces
contractivity of the passage map, which partitions trajectories into
fast (direct-transit) and slow (reinjection) populations with no
stable intermediate pathway. Quantitative analysis across 59 parameter
combinations reveals that for $\rho$ sufficiently above the onset of
chaos the gap width $\Delta t_{\mathrm{gap}} \approx 0.68 \pm 0.01$ is
approximately invariant, constant to within 3.3\% despite substantial
parameter variation ($\sigma \in [8,14]$, $\rho \in [24,38]$, $\beta
\in [2.0, 3.3]$). Near the Hopf bifurcation, the gap width exhibits a
systematic correlation with $\rho$ ($R^2 \approx 0.45$), consistent
with the gradual dissolution of the bimodal lobe structure. Neither
$\beta$, $\lambda_u$, nor $\nu$ explains the gap width ($R^2 < 0.03$
for all models tested away from the bifurcation), indicating that both
gap boundaries are set by the global geometry of the branched manifold
rather than by local dynamics at the origin.

\emph{Experimental interpretation.} The auxiliary variable $v_n(t)$
admits interpretation as an integrated heat flux anomaly in the
context of Rayleigh-B\'enard convection. While the precise
experimental implementation faces practical challenges due to the
Lorenz system's status as a truncated model, this connection suggests
that the invariant construction encodes physically meaningful
information beyond its formal mathematical structure.

The framework established here opens several avenues for future
investigation. Whether analogous geometric constraints govern the
scaling behavior for other sensor pairs from different classes remains
to be explored. The complete structure of the 192-element family
obtained by allowing independent component signs has not been
investigated. The parametric sweep establishes that the exponent $n$
scales as $\beta^{0.85}$ and $\rho^{-0.69}$ individually, steeper than
the naive $\sqrt{\beta/\rho}$ expectation from fixed-point
coordinates; deriving these exponents from the structure of the
Poincar\'e return map near $C_\pm$ constitutes the most concrete open
problem in the scaling analysis. The non-monotonic
$\sigma$-dependence, with a minimum near $\sigma = 10$ and elevated
values at both small and large $\sigma$, suggests competing mechanisms
whose interplay remains to be elucidated. The precise characterization
of the combinatorial robustness of Class~III invariants, including the
determination of the maximal perturbation amplitude compatible with
symbolic itinerary preservation and its possible connection to
quantized topological invariants of the branched manifold, constitutes
a further open problem. The analytical derivation of the gap width
$\Delta t_{\mathrm{gap}} \approx 0.68$ from the global geometry of the
branched manifold offers a concrete and now well-characterized target
for future work. Most significantly, the question of whether analogous
history-dependent invariants exist in other dissipative chaotic
systems suggests that the Lorenz construction may represent a specific
instance of a more general phenomenon. The systematic methodology
developed here, comprising permutation-based enumeration with
statistical characterization, topological precursor analysis,
dynamical branch identification, and parametric robustness
verification, provides a template for such investigations.

\section*{Acknowledgments}

This research was conducted independently at the Departamento de
F\'{\i}sica, Universidad de Chile, without external funding.

\section*{Declaration of competing interest}

The author declares that he has no known competing financial interests
or personal relationships that could have appeared to influence the
work reported in this paper.

\section*{Declaration of generative AI and AI-assisted technologies in the writing process}

During the preparation of this work, the author used an AI language
model to improve the readability and internal consistency of the text,
and for assistance with background research. After using this tool,
the author reviewed and edited the content as needed and takes full
responsibility for the content of the publication.

\section*{Data availability}

The numerical data supporting the findings of this study were
generated using custom scripts in Mathematica (for symbolic analysis
and high-precision integration) and Julia (for large-scale statistical
sampling). The integration employed adaptive step-size control with
extended-precision arithmetic (80 decimal digits) for validation
calculations and standard double precision for statistical ensemble
generation. The robust validation analysis comprised 59 parameter
combinations with 5 realizations per point and 100 bootstrap samples
per realization, totaling approximately 1.5 million precursor
events. All computational scripts and representative datasets are
available from the corresponding author upon reasonable request.


\end{document}